\documentclass[letterpaper]{emulateapj}
\usepackage{apjfonts}

\newcommand{\fig}[1]{\mbox{Fig.~\ref{#1}}}
\newcommand{\figs}[2]{\mbox{Figures~\ref{#1}~\&~\ref{#2}}}
\newcommand{\tab}[1]{\mbox{Table~\ref{#1}}}
\newcommand{\sect}[1]{\mbox{\S~\ref{#1}}}
\newcommand{\append}[1]{\mbox{Appendix~\ref{#1}}}

\newcommand{\Ha}{\mbox{H$\alpha$}}
\newcommand{\Hb}{\mbox{H$\beta$}}
\newcommand{\Hd}{\mbox{H$\delta$}}

\newcommand{\Hydrogen}{\mbox{H}}
\newcommand{\HI}{\mbox{H\,{\sc i}}}
\newcommand{\HII}{\mbox{H\,{\sc ii}}}
\newcommand{\fNII}{\mbox{[N\,{\sc ii}]}}
\newcommand{\fOI}{\mbox{[O\,{\sc i}]}}
\newcommand{\fOII}{\mbox{[O\,{\sc ii}]}}
\newcommand{\fOIII}{\mbox{[O\,{\sc iii}]}}
\newcommand{\fSII}{\mbox{[S\,{\sc ii}]}}
\newcommand{\fSIII}{\mbox{[S\,{\sc iii}]}}

\newcommand{\snr}{\mbox{S/N}}
\newcommand{\lam}{\mbox{$\lambda$}}

\newcommand{\LCDM}{\mbox{$\Lambda$CDM}}
\newcommand{\EBV}{\mbox{$E(\bv)$}}
\newcommand{\nbody}{\mbox{$N$-body}}

\newcommand{\s}{\mbox{s}}
\newcommand{\yr}{\mbox{yr}}
\newcommand{\Myr}{\mbox{Myr}}
\newcommand{\Gyr}{\mbox{Gyr}}

\newcommand{\ang}{\mbox{\AA}}
\newcommand{\cm}{\mbox{cm}}
\newcommand{\km}{\mbox{km}}
\newcommand{\pc}{\mbox{pc}}
\newcommand{\kpc}{\mbox{kpc}}
\newcommand{\Mpc}{\mbox{Mpc}}

\newcommand{\erg}{\mbox{erg}}

\newcommand{\scM}{\mathcal{M}}

\newcommand{\gi}{\mbox{$g-i$}}
\newcommand{\VI}{\mbox{$V-I$}}

\newcommand{\TwodFGRS}{\mbox{2dFGRS}}
\newcommand{\CFRS}{\mbox{CFRS}}
\newcommand{\CNOC}{\mbox{CNOC}}
\newcommand{\SDSS}{\mbox{SDSS}}
\newcommand{\HST}{\mbox{\emph{HST}}}

\newcommand{\Stromgren}{\mbox{Str\"{o}mgren}}

\newcommand{\vlos}{\mbox{$v_{\rm los}$}}

\newcommand{\EBVr}{\mbox{$E_*/E_{\rm g}$}}

\defcitealias{dre99}{D99}
\defcitealias{kod01}{K01}
\defcitealias{mlf00}{MLF00}
\defcitealias{sat06}{Paper~I}

\newcommand{\clustercenter}{$9^{\rm h}42^{\rm m}56\fs2$,
$46^\circ59'12\farcs0$ (J2000)}
\newcommand{\clustercenterXray}{$9^{\rm h}43^{\rm m}03\fs3$,
$46^\circ59'26\farcs5$ (J2000)}
\slugcomment{To appear in Astrophysical Journal}
\shorttitle{Infalling Faint \fOII\ Emitters in A851. II.}
\shortauthors{Sato \& Martin}

\begin{document}

\title{Infalling Faint \fOII\ Emitters in \object{Abell 851}.
II.~Environment, Kinematics, \& Star Formation History}
\author{Taro Sato\altaffilmark{1} and Crystal
L. Martin\altaffilmark{2,3,4}} \affil{Department of Physics,
University of California, Santa Barbara, CA 93106-9530}
\altaffiltext{1}{taro@physics.ucsb.edu}
\altaffiltext{2}{cmartin@physics.ucsb.edu}
\altaffiltext{3}{Packard Fellow}
\altaffiltext{4}{Alfred P. Sloan Foundation Fellow}

\begin{abstract}
We report on the local environment, radial velocity, and star
formation history of \fOII-selected emission-line objects in the
$z\approx 0.4$ galaxy cluster \object{Abell 851}, using optical
spectra obtained with Keck~I/LRIS and Keck~II/DEIMOS.  A large
fraction ($\approx 55\%$) of cluster \fOII\ emitters show strong
Balmer absorption lines ($\ga 4~\ang$ in equivalent width at \Hd).
These e(a)-type spectra have been attributed to dusty starburst
galaxies by \citet{pog00}, an interpretation supported by our
Balmer-decrement--derived reddening measurements, which show a high
frequency of very reddened [$\EBV \ga 0.5$] \fOII\ emitters.  Our
spectral modeling requires starburst ages $\la 1~\Gyr$, which is
shorter than the cluster crossing timescale.  We argue that this
starburst phase occurs during cluster infall based on the radial
velocity distribution of the cluster \fOII\ emitters, which present a
deficit of systems near the cluster systemic velocity compared to a
virialized population (or a backsplash population).  The spatial
segregation of some redshifted and blueshifted groups strongly
indicates that the accretion was recent.  Throughout the cluster, the
fraction in \fOII-emitting galaxies is a strong function of the local
galaxy density, being suppressed in dense environments.  Our analysis
supports previous suggestions that dusty starburst galaxies arise at
the expense of continuously star-forming gas-rich spiral galaxies
[i.e., the e(c)-type; \citet{dre99}].  In addition, we describe a
fainter \fOII-emitting population, comprised largely of dwarf galaxies
\citep{mlf00} and find an even stronger suppression of \fOII\ emission
in high density environments among this subsample, indicative of more
effective destruction by harassment and/or gas-stripping.  Comparison
to previous morphological studies, limited to the central region of
\object{Abell 851}, suggests that galaxy-galaxy interactions may
trigger the starbursts.  The high e(a) galaxy fraction in
\object{Abell 851} compared to that in the field, however, suggests
that some cluster-specific mechanism, likely related to the dynamical
assembly of the cluster, also contributes to the high number of
starbursts.
\end{abstract}


%\received{}
%\accepted{}

\section{Introduction}

Local environment has a strong effect on the star-forming activity and
morphology of galaxies---the observation first recognized in the cores
of rich clusters \citep{dre80, but84}.  There is mounting evidence in
support for the Butcher-Oemler effect \citep{but84} as a cosmic
process, not specific to galaxy clusters, in which the
field-to-cluster transformation of galaxies occurs upon the accretion
of field galaxies to cluster central regions \citep{dre94, cou94,
abr96, ell01, tra04}.  In fact, recent studies of cluster galaxies
have shown that mechanisms responsible for their transformation must
be in effect well beyond cluster virial radii, well before the
galaxies at infall reach the central regions of clusters
\citep[e.g.,][]{bal97, tre03, rin05, mor05, pim06}.  Since cluster
environment samples a wide range in local galaxy density and relative
velocity of cluster members, a dominant physical process driving
galaxy evolution may vary over an infall period.  Thus most recent
efforts in cluster galaxy surveys have focused on studying galaxies in
infall regions where the preprocessing of cluster galaxies is already
in effect.

Although the studies spanning a few decades have established such
empirical relations as the morphology-density relation \citep{dre80}
and the Butcher-Oemler effect \citep{but84} for giant luminous cluster
galaxies, the underlying physical processes that are responsible for
the transformation of their properties and therefore the
aforementioned empirical relations have yet to be well constrained to
date.  The sign of the differential evolution between dwarf and giant
galaxies is even more obscure, since few wide-field surveys have
sampled faint galaxies at high redshift.  Dwarf and large spiral
galaxies are expected to respond differently to cluster tidal fields
due to their different mass concentrations \citep{moo99}.  In
predicting a steep halo mass function, the currently standard
hierarchical galaxy formation models \citep[e.g.,][]{kau93, col94}
produce more dwarf galaxies than actually observed
\citep[e.g.,][]{kly99, som02}.  Galaxy clusters host dense
environments that are hostile to fragile dwarf galaxies; therefore the
apparent faint-end steepening of cluster galaxy luminosity function
presents a challenge to dwarf formation models \citep[see, e.g.,][and
references therein]{pop05}.

Redshift surveys of local universe and \nbody\ simulations show that
large-scale filaments are a natural product of gravitationally driven
structure formation \citep[e.g.,][]{bon96, rin01, pim04}.  Galaxy
clusters are expected to form at the intersections of filamentary
structures, yet there has been few studies on the direct detection of
filaments extending out from isolated clusters at high redshifts
\citep[e.g.,][]{ebe04}.  A rich, moderately distant ($z \simeq
0.4069$) galaxy cluster \object{Abell 851} [\object{CL0939+4713};
\clustercenter] offers evidence of filaments, observed via the
photometric redshift technique \citep[][hereafter
\citetalias{kod01}]{kod01}.  Since few clusters at this epoch have
been studied well beyond their virial radii, it remains unclear
whether the observed structures are actually filaments of cosmological
origin or are related to on-going merger event \citep{def03}.  In
either case, \object{Abell 851} presents a unique opportunity to study
cluster galaxy evolution in a dynamically active environment on the
large scales.

Before the identification of the filaments, \object{Abell 851} had
already drawn attention due to the extreme richness
\citep[e.g.,][]{gun86} and high abundance of blue, star-forming
galaxies; the cluster contains a larger fraction in young
emission-line galaxies compared to clusters at similar redshift
\citep{lot03}.  A spectroscopic survey of $r \la 22.5$ galaxies
revealed a high fraction in post-starburst galaxies, whose spectra are
dominated by Balmer absorption lines of intermediate-age stars
\citep[][hereafter \citetalias{dre99}]{dre99}.  The timescales
associated with such galaxies implied the suppression of star-forming
activity within the past $\sim 1.5~\Gyr$.  Since the cluster crossing
time for infalling galaxies is at least a few \Gyr\ \citep{tre03} and
definitely longer than the post-starburst timescales, the high
abundance of post-starburst galaxies suggested the significance of
cluster-induced starburst in the transformation of cluster galaxies.
Although the scarcity of star-forming activity in very dense
environments within the central regions of clusters has been
universally observed and confirmed, there has been a debate as to the
existence of enhanced starburst phase \emph{before} the suppression of
star-forming activity of infalling cluster galaxies kicks in
\citep{dre99, bal99, tra03, tra04}.  In post-starburst galaxies, we
see remnants of starbursts, but detecting the signs of starbursts is
generally complicated by the need for constraining the star formation
history (SFH) of galaxies---the process requiring more than a simple
diagnostics of star-forming activity.

\citet[][hereafter \citetalias{mlf00}]{mlf00} carried out narrowband
and broadband photometry of \object{Abell 851} using Mayall 4-meter
telescope at the Kitt Peak National Observatory, to detect faint
emission-line galaxies using their redshifted $\fOII\lam3727$ emission
line.  The narrowband search yielded $371$ emission-line cluster
member candidates.  The completeness limit of the \citetalias{mlf00}
survey was $g \simeq 24.5$, about ten times deeper than the
\citetalias{dre99} spectroscopic survey.  The locus in the $(\gi)$-$i$
color-magnitude diagram and their small isophotal area suggested their
dwarf identity for a significant fraction of the \fOII\ emission
candidates.  In a spectroscopic follow-up program of the
\citetalias{mlf00} survey, we have obtained an extensive set of
spectra of the \fOII-emitting candidates using the Keck~I Low
Resolution Imaging Spectrometer \citep[LRIS;][]{oke95} and Keck~II
Deep Imaging Multi-Object Spectrograph \citep[DEIMOS;][]{fab03}.

As the second in a series of papers drawing on the Keck spectroscopic
sample of the \fOII\ emitters, we present the analysis of environment,
kinematics properties, and SFH of infalling faint star-forming
galaxies in a highly dynamically active galaxy cluster at $z\simeq
0.4$.  The observations, data, and the spectroscopic confirmation of
\citetalias{mlf00} \fOII\ emitters have been reported in
\citet[][hereafter \citetalias{sat06}]{sat06}, to which readers are
referred for the detailed information on the basic characteristics of
our spectroscopic sample.  This paper starts with the sections
entirely dedicated to the effect of local environment
(\sect{sec.EffectsOfLocalEnvironment}) and the kinematic properties
(\sect{sec.KinematicsOfTheClusterOIIEmitters}) of cluster \fOII\
emitters.  In \sect{sec.ClassificationBySpectralType} we present the
cluster \fOII\ emitters in terms of their spectral types derived from
the \fOII\ emission and \Hd\ absorption equivalent widths, which will
be analyzed in view of of photoionization and stellar population
synthesis models in \sect{sec.SFHOfClusterOIIEmitters}.  We then
discuss (\sect{sec.Discussion}) and summarize (\sect{sec.Summary}) our
results.

Throughout this paper, we use the standard \LCDM\ cosmology,
$(\Omega_m, \Omega_\Lambda) = (0.3, 0.7)$, with $H_0 =
70~\km~\s^{-1}~\Mpc^{-1}$.  At the distance of \object{Abell 851} ($z
= 0.4069$), $1''$ on the sky corresponds to a projected physical
distance of $5.43~\kpc$, and the lookback time to the galaxy cluster
is $4.3~\Gyr$.

\section{Effects of Local Environment}
\label{sec.EffectsOfLocalEnvironment}

The star-forming property of a galaxy has been shown to depend
strongly on its local environment, often quantified by the projected
number density of galaxies near the apparent position of the galaxy in
the sky.  For \object{Abell 851} in particular, \citetalias{kod01}
have reported a so-called threshold effect---galaxy \VI\ colors
abruptly become redder above a threshold density, $\sim 50~{\rm
galaxies}~\Mpc^{-2}$ (in the cosmology assumed in this paper), which
roughly corresponds to the typical local galaxy density within the
substructures in surrounding filaments.\footnote{We caution that
local galaxy densities computed in such a method used in our study may
not be directly compared to those in other surveys; a projected
density estimate derived from a nearest-neighbor approach is highly
sensitive to the specific ways in which foreground and background
galaxies are removed from a cluster sample.}  Robust determinations
of cluster memberships are essential for quantifying local environment
of cluster galaxies.  We obtained the photometry and photometric
redshifts of the objects studied in \citetalias{kod01}, kindly
provided by T.~Kodama.  The \citetalias{kod01} survey limiting
magnitude was roughly comparable to our survey: $I({\rm AB}) \simeq
23.4$ (\citetalias{kod01}) and $i \simeq 23$ (\citetalias{mlf00});
this is quite deep, as \citetalias{kod01} estimates $M_V^*$ to be
about $I=19.4$ in this cluster.

\subsection{Local Projected Density of Galaxies}
\label{sec.LocalProjectedDensityOfGalaxies}

The local environment of a galaxy is quantified by a projected surface
galaxy number density, $\Sigma$, using ten nearest cluster members
defined by \citetalias{kod01} via the photometric redshift method.
The value is computed for each galaxy in our survey as follows:
(1)~Using the RA and DEC of a galaxy, we find ten \citetalias{kod01}
confirmed cluster members nearest to that position; (2)~compute the
distance between the farthest and closest members to that reference
coordinates in the ten objects just found; and (3)~take that as the
radius of a circular projected area by which the number of galaxies
within the area (i.e., always $10$) is divided.  The only difference
from the conventional method is that our ten-nearest neighbor density
may or may not include the galaxy for which the value is computed.
This is because explicit spatial cross-correlation was not made
between the \citetalias{mlf00} and \citetalias{kod01} objects, but the
deviation does not significantly alter the conclusions drawn from our
study.  \citetalias{kod01} estimated the field contamination to the
projected density of cluster galaxies in their survey to be $2.29 \pm
0.06~{\rm arcmin}^{-2}$.  We statistically subtracted this from our
computation of local galaxy density described above.  In the
nearest-neighbor approach, the effect of Poissonian uncertainty in the
expected number of field galaxies becomes dominant for smaller
$\Sigma$, since the area enclosing ten cluster members contains many
more non-member objects.  A $1\sigma$ Poisson uncertainty in the field
contamination becomes comparable to the local density estimate for a
cluster member at $\log\Sigma \simeq 0.3$.  The systematic removal of
field contamination yields unphysical value of $\Sigma < 0~\Mpc^{-2}$
for some galaxies.  Where required, we generally assigned $\log\Sigma
= -2$ as the upper limit for such galaxies.

\subsection{\fOII\ Emitter Fraction}

We start with a brief summary of the basic properties of the cluster
\fOII-emitting sample presented in \citetalias{sat06}.  Our cluster
\fOII\ emitters for this spectroscopic follow-up were drawn from $371$
narrowband-selected objects with their candidate \fOII\ emission
detected at $\snr>3$ level in \citetalias{mlf00}.  The narrowband
survey was roughly complete to the observed equivalent width of $\la
-11~\ang$ at $m_{5129}({\rm AB}) \simeq 24$.\footnote{An equivalent
width of an emission line is designated a negative sign; see
\citetalias{sat06} for the sign and other notational conventions used
in this paper. } The preference was given to the objects with bluer
\gi\ colors (i.e., $\gi < 1$), so the sample is more complete for blue
objects.  Among the $212$ objects selected for spectroscopic
follow-up, $111$ actually showed \fOII\ emission and had redshifts
within the range $z = 0.395$--$0.420$, constituting our cluster
\fOII-emitting sample.  This final sample spans a wide range in
luminosity, $18.5 \la i \la 25.5$, where $i \simeq 23$ corresponds to
the completeness limit of the \citetalias{mlf00} narrowband survey.

\begin{figure}
\plotone{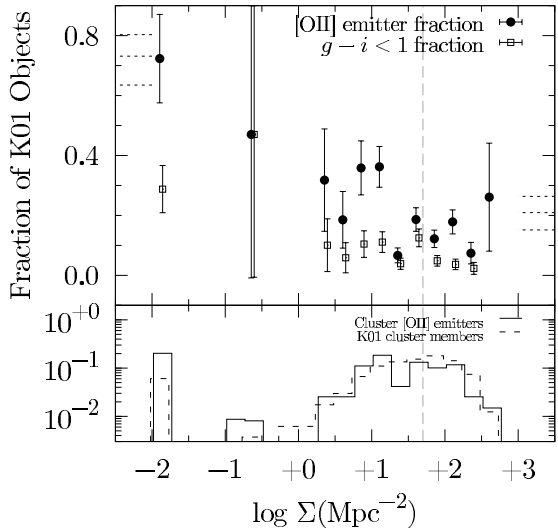}
\caption{ (\emph{Top}) The \fOII\ emitter fractions (filled circles)
as a function of local galaxy density $\Sigma$.  The fractions of
\fOII\ emitters in $\gi < 1$ galaxies (open squares) are also shown.
The fractions are computed from the number of
spectroscopically-confirmed \fOII\ emitters (multiplied by appropriate
correction factors given in \citetalias{sat06} to account for the
selection bias) normalized to the number of \citetalias{kod01} cluster
members at each density bin.  Dotted horizontal lines indicate the
best estimates and $1\sigma$ uncertainties for the fractions of
galaxies with $W_0(\fOII) < -10~\ang$ in cluster and field samples at
$z \simeq 0.4$ computed by \citet{nak05}.  (\emph{Bottom}) Fractional
abundance as a function of local galaxy density for the
spectroscopically-confirmed cluster \fOII\ emitters corrected for the
target selection bias ($N\simeq175$; solid histogram) and the
\citetalias{kod01} cluster members ($N=808$; dashed histogram).  The
fractional abundance is normalized to the total number of objects
($N$) within each sample to facilitate the comparison of density
distributions.  The vertical gray dashed line indicates the threshold
density, $\log\Sigma \simeq 1.7$, over which \VI\ colors of galaxies
become redder as reported by \citetalias{kod01}.  }
\label{fig.EmissFracVsLogSigma}
\end{figure}

\fig{fig.EmissFracVsLogSigma} shows the distributions of cluster
\fOII\ emitters and the \citetalias{kod01} cluster members within the
survey area of \citetalias{mlf00}.  Note that the histogram for the
spectroscopically-confirmed cluster \fOII\ emitters are corrected for
the target selection bias by the \gi\ color, i.e., we multiply the
number actually observed by the the correction factor for each \gi\
color subsample; see \citetalias{sat06} for the details.  A
Kolmogorov-Smirnov (K-S) test of the distributions of local galaxy
density for the spectroscopically-confirmed cluster \fOII\ emitters
and more general, photo-$z$--selected cluster population from
\citetalias{kod01} yields a very small probability ($\ll 1\%$) that
they share a common parent distribution.  This indicates the
\fOII-emitting population resides in different environments than the
general cluster galaxy population.  A relatively higher fraction of
cluster \fOII\ emitters is found at low densities.  The emission-line
galaxy fraction has been shown observationally to depend strongly on
local environments, where a smaller fraction is observed in dense
environments \citep[e.g.,][]{bal04b}.  Furthermore, such an
emission-line galaxy fraction appears to evolve with redshift
\citep{kod04, nak05}.

At the lowest density bin, the fraction in \fOII\ emitters matches
well the field \fOII\ emitter fraction at $z\sim0.4$ of $\sim 70\%$
estimated by \citet{nak05} from their sample drawn from the Canadian
Network for Observational Cosmology~1 (\CNOC1) cluster survey; the
\fOII\ equivalent width cut for their \fOII-emitting population is
similar to ours [$W_0(\fOII) < -10~\ang$].  As discussed in
\citetalias{sat06}, \citetalias{mlf00} could have detected only $\sim
75\%$ of virialized cluster \fOII\ emitters with their narrowband
filter, assuming a Gaussian velocity distribution for virialized
cluster members.  In \fig{fig.EmissFracVsLogSigma}, we did not
statistically correct for this population which could have been missed
by the \citetalias{mlf00} narrowband survey.  Although we do not
expect the \citetalias{mlf00} survey area ($4.4~\Mpc \times 4.4~\Mpc$)
to be large enough for sampling a true field population and applied a
crude correction for target selection bias, the sampling at
low-density region appears to provide us with a reasonable
\fOII-emitting sample in field-like environments, since their observed
fraction is consistent with the independent measurement of field
\fOII\ emitter fraction.

\begin{figure}
\plotone{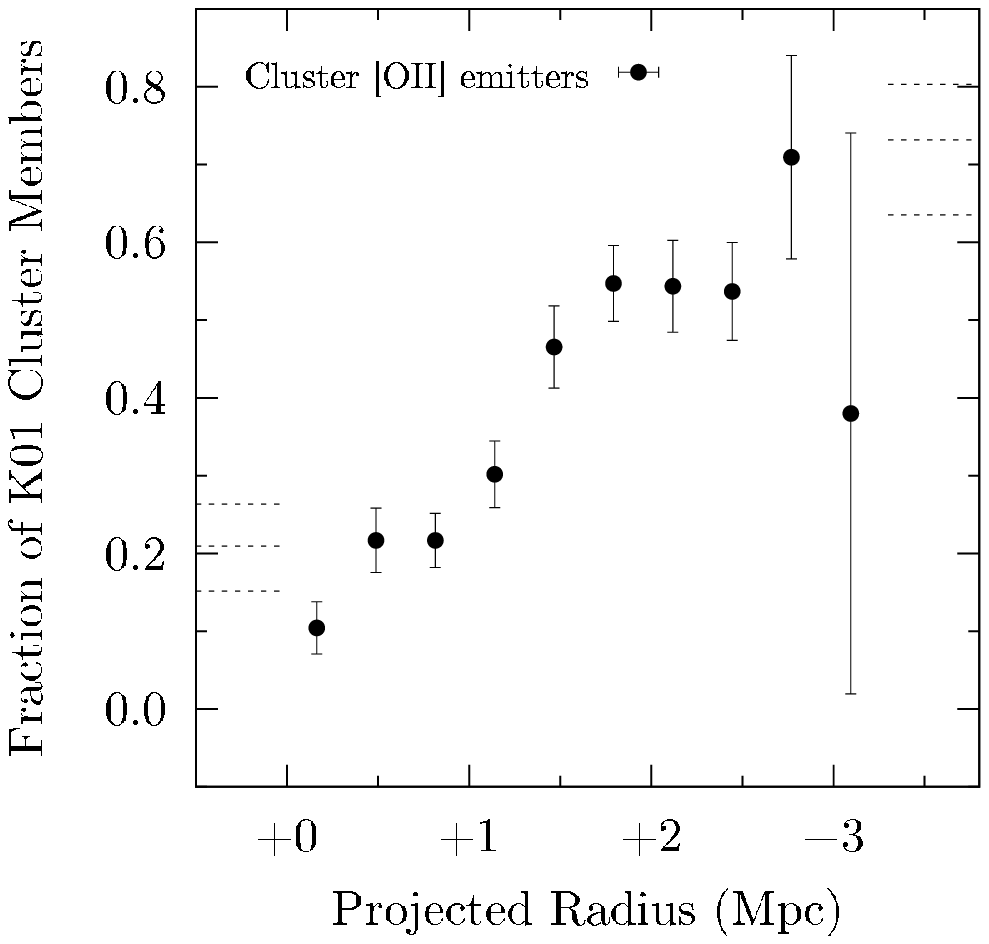}
\caption{ The \fOII\ emitter fraction of \citetalias{kod01} photo-$z$
cluster members as a function of projected radius from the cluster
center.  The center of the cluster was taken to be at
\clustercenterXray, corresponding to the brightest point in X-ray
surface brightness from \citet{def03}.  Dotted horizontal lines
indicate the best estimates and $1\sigma$ uncertainties for the
fractions of galaxies with $W_0(\fOII) < -10~\ang$ in cluster and
field samples at $z \simeq 0.4$ computed by \citet{nak05}.  At each
radial bin, the fraction is corrected for \citetalias{mlf00}
emission-line candidates not spectroscopically observed, in addition
to the usual target selection correction by \gi\ colors
\citepalias{sat06}.  }
\label{fig.O2FracVsR}
\end{figure}

The \fOII\ emitter fraction in the densest environment in our sample,
$\sim 20\%$, is also roughly consistent with the observed \fOII\
emitter fraction in $z=0.4$ clusters \citep{nak05}.  In relation to
the \citetalias{kod01} threshold density at which the \VI\ colors of
cluster galaxies have been shown to become red quite abruptly, the
presence of \fOII\ emitters appears already suppressed at slightly
lower densities, and their fraction remains roughly constant at the
lowest level toward denser environments.  \fig{fig.O2FracVsR} shows
the \fOII\ emitter fraction as a function of projected radius from the
cluster center.  The radial trend is expected in view of
\fig{fig.EmissFracVsLogSigma}, since the local galaxy density is
generally anti-correlated with the cluster radius.

\begin{figure}
\plotone{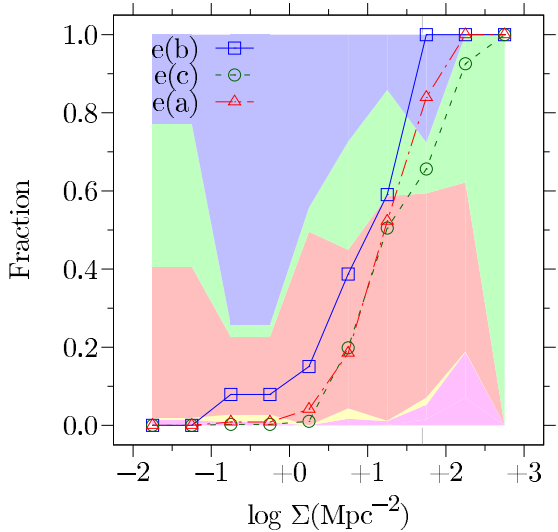}
\caption{ Cumulative local density distributions of the cluster \fOII\
emitters with various \citetalias{dre99} spectral types: e(b) (open
squares), e(c) (open circles), and e(a) (open triangles).  The
distribution is normalized to zero at the field value (i.e., at the
lowest density bin) for each spectral type.  The samples are binned
with constant $\log{\Sigma}=0.5$ intervals, where the symbols in the
plot indicates their central values.  The spectral types are assigned
using the best estimates as well as the uncertainties in \fOII\ and
\Hd\ measurements (see
\sect{sec.DistributionOfOIIEmittersBySpectralClass} for detail).  The
fractions are corrected for the target selection bias by appropriate
correction factors \citepalias{sat06}.  In the electronic edition, the
relative abundances of the \fOII\ cluster emitters in different
\citetalias{dre99} spectral types are also indicated by the filled
curves: from the top, they are e(b) (blue), e(c) (green), e(a) (red),
k (yellow), and k+a/a+k (magenta).  The colors are half-toned for
better contrast.  [\emph{See the electronic edition of the paper for a
color version of this figure.}] }
\label{fig.SCVsLogSigma}
\end{figure}

The distribution of $\gi<1$ galaxies among the \fOII\ emitters shows a
modest peak near the \citetalias{kod01} threshold density.  It is hard
to see whether this peaking is directly related to the threshold
effect, but it could indicate the presence of starbursts as galaxies
enter local environments typical of cluster substructures.  In
\fig{fig.SCVsLogSigma}, we see the cumulative local density
distributions of the cluster \fOII\ emitters with various
\citetalias{dre99} spectral types.  According to $\chi^2$ tests, the
cumulative distributions among e(b), e(c), and e(a) types are
different at very high confidence levels ($\gg 99\%$).  The details of
the \citetalias{dre99} classification scheme will be discussed in
\sect{sec.DistributionOfOIIEmittersBySpectralClass}, but for the
current discussion, we note that e(b), e(a), and k+a/a+k galaxies are
all associated with starbursts, whereas e(c) and k galaxies are not.
\fig{fig.SCVsLogSigma} may provide additional evidence for a higher
incidence of starbursts in the local environments near the
\citetalias{kod01} threshold of $\log\Sigma \simeq 1.7$.  We also note
a striking drop in the \fOII\ emitter fraction at $\log\Sigma \approx
1.3$, not too far from the \citetalias{kod01} threshold density.  Although a very strong suppression of star-forming activity near the
\citetalias{kod01} threshold could be real, the \fOII\ emitter and
$\gi<1$ galaxy fraction are already well below the field values at
$\log\Sigma \approx 0.5$; therefore, some mechanism of quenching star
formation could be in effect below the \citetalias{kod01} threshold
density as well.

\subsection{\fOII\ \& \Ha\ Equivalent Width Distribution}

\begin{figure}
\plotone{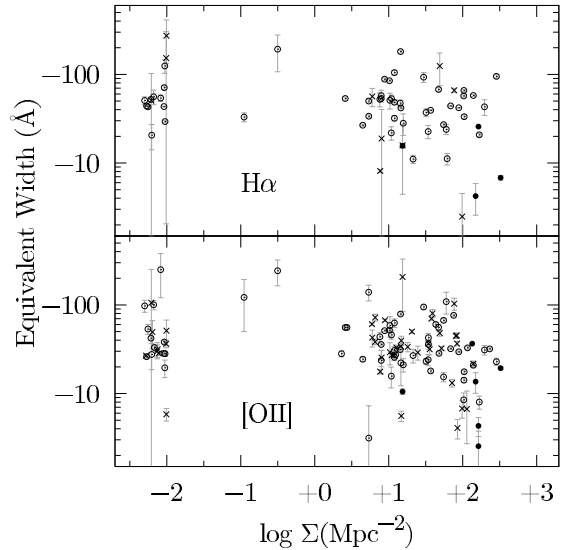}
\caption{ Rest-frame $\fOII\lam3727$ (\emph{bottom}) and \Ha\
(\emph{top}) equivalent widths of cluster \fOII\ emitters as a
function of their local galaxy density.  Open and filled circles are
\HII\ region-like galaxies and AGN-like objects, respectively; crosses
are the objects whose classification by the line ratio diagnostics
could not be carried out.  See \citetalias{sat06} for the method used
for these identifications.  For the data points with $\log\Sigma =
-2$, small negative random numbers are added to their densities to
avoid excessive overlapping.  }
\label{fig.W0eVsLogSigma}
\end{figure}

Although the star-forming galaxy fraction strongly depends on local
environments, in previous studies the amount or strength of
star-forming activity within each galaxy, probed via such
emission-line diagnostics as \Ha\ equivalent width, does not appear to
depend strongly on local environments \citep{bal04a, bal04b, kod04,
rin05}.  In \fig{fig.W0eVsLogSigma} we see how our cluster
\fOII-emitting sample compares.  For the wide dynamic range in local
galaxy densities, almost five decades in $\Sigma$, the strength of
\fOII\ and \Ha\ equivalent widths shows only a mild decrease at high
densities.  The lack of dependence on local galaxy densities becomes
more remarkable in view of \fig{fig.CoaddGridsIMagLogSigma}, where
luminous \fOII\ emitters---many of which are found to be AGN-like
\citepalias{sat06}---have weaker \fOII\ and \Ha\ equivalent widths.
The high abundance of luminous AGN-like \fOII\ emitters at high
densities significantly accounts for the apparent trend.

Relative independence of the strength of emission equivalent widths of
star-forming galaxies on their local galaxy density, combined with the
strong trend of decreasing \fOII\ emitter fraction with local galaxy
density, has given rise to an interpretation that the mechanisms that
halt star-forming activity in these galaxies either work on very short
timescales and/or are sufficiently rare that we are not seeing an
abundance of galaxies in transition from actively star-forming to
passively-evolving states.  Starbursts followed by quenching of
star-formation offers an attractive solution in this regard.  A
fraction of total galaxies in the starburst or post-starburst phase
could offer an insight, yet the existing surveys have shown
considerable disagreements as to the frequency of starburst and/or
post-starburst galaxies \citep{dre99, bal99, tra03, tra04}.  Even with
a robust estimate of post-starburst fraction, the role of
post-starburst phase in galaxy evolution is still obscure.
Nonetheless our cluster \fOII-emitting sample offers no contradicting
evidence to either of the above interpretations.

\subsection{Luminosity Distribution}
\label{sec.LuminosityDistribution}

\begin{figure}
\plotone{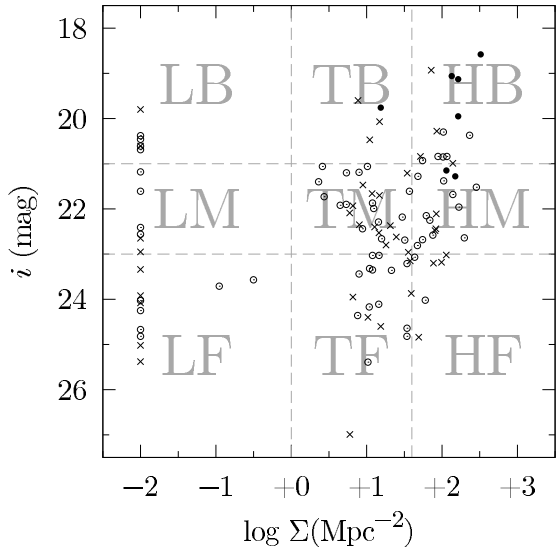}
\caption{ The $i$ magnitudes of cluster \fOII\ emitters as a function
of their local galaxy density.  Symbols are as in
\fig{fig.W0eVsLogSigma}.  Grids of the $i$-$\Sigma$ plane defines the
subsamples of cluster \fOII\ emitters to be coadded
(\sect{sec.SFHByCompositeSpectra}).  The magnitude cuts are made at $i
= 21$ and $23$, and the density cuts are at $\log{\Sigma} = 0$ and
$1.6$.  Objects are divided into nine subsamples by their
$\log\Sigma$: low-density (L), transitional (T), and high-density (H)
environments; and by their luminosity: bright (B), medium (M), and
faint (F) in term of $i$ magnitude. }
\label{fig.CoaddGridsIMagLogSigma}
\end{figure}

In \fig{fig.CoaddGridsIMagLogSigma} we see the distribution of cluster
\fOII\ emitters in terms of their luminosity and local environment.
Interestingly, it shows that among our cluster \fOII\ emitters
AGN-like galaxies are luminous and mostly reside in dense
environments.  However, we must caution that within the intermediate
luminosity range there are a number of galaxies for which we could not
carry out line-ratio diagnostics for AGN activity \citepalias{sat06},
indicated by crosses in the figure.  Also, none of the cluster \fOII\
emitters are spatially coincident with the \emph{XMM-Newton} X-ray
point sources detected at the level of $L_X({\rm bol}) \ga 1.6 \times
10^{43}~\erg~\s^{-1}$ by \citet{def03}.  The X-ray observation does
not exclude the possibility of these galaxies hosting nuclear
activities, however.  In \fig{fig.GIClrMag} we see the AGN-like
galaxies in our sample have $\gi > 1$, putting them in the reddest end
of the distribution of \SDSS\ red AGNs, which generally appear X-ray
weak as well \citep{bra04}.  If the observed trend of AGN-like galaxy
distribution is real, it would point to the significant role of
optically-luminous AGN-like galaxies in dense environments, whose
\fOII\ emission equivalent widths are generally smaller than those of
more general star-forming galaxy population \citepalias{sat06}.  As
mentioned in \citetalias{sat06}, if AGN feedback acts to suppress star
formation in luminous (or massive) galaxies, that offers a possible
explanation for the \emph{downsizing effect}, which refers to
observational evidence for the redshift evolution of characteristic
galaxy luminosity above which star-forming activities are suppressed
\citep{cow96}.

It is also notable that we see relatively few low-luminosity galaxies
in dense environments, in the region marked as HF in
\fig{fig.CoaddGridsIMagLogSigma}.  Although the range $i>23$ is below
the completeness limit of narrowband survey, the depletion of faint
galaxies at high densities is remarkable because of the abundance of
galaxies with similar luminosities in lower-density environments.  The
relative depletion of faint galaxies in high-density environments here
simply means the lack of star-forming population, and can result from
suppressed star-forming activity in faint galaxies, which simply fade
out of our survey limit, or their true absence, possibly due to the
destruction of low surface brightness dwarf galaxies within a strong
cluster gravitational potential or by a series of impulsive
gravitational encounters with other galaxies---so-called galaxy
harassment \citep{moo99}.  The $i$-band magnitude of our \fOII\
emitters samples their rest-frame $V$-band light, which itself depends
on the amount of on-going star-forming activity.  Since our survey is
insensitive to the non--star-forming population, we cannot rigorously
exclude or favor either of the two possibilities.

In \citetalias{sat06} we showed the strength of \fOII\ equivalent
width shows only a mild decreasing trend with luminosity.  In dense
environments, the absence of strong \fOII\ emitters is partially a
consequence of the lack of faint \fOII\ emitters and the presence of
optically-luminous AGN-like galaxies with moderate to weak \fOII\
equivalent widths.  This environmental trend of \fOII-emitting
population makes stronger the claim that the amount of star-forming
activity in star-forming galaxies does not vary much in different
local environments, while the star-forming galaxy fraction is a strong
function of local environment.  The luminosity segregation of galaxy
populations in dense environments further suggests the importance of
differential evolution of galaxies based on their intrinsic
properties.

\section{Kinematics of the Cluster \fOII\ Emitters}
\label{sec.KinematicsOfTheClusterOIIEmitters}

Studying the connections between galaxies and their host environments
in galaxy clusters helps to understand the transformation of galaxies
in cosmological context.  Cluster galaxy populations in relation to
projected cluster-centric radii and galaxy local number densities are
relatively easier to quantify and therefore have been well studied
\citep[e.g.,][]{oem74, mel77, tre03}.  The relations between cluster
galaxy populations and their kinematics, however, are nontrivial since
cluster galaxies are not necessarily in dynamical equilibrium on the
scale of an entire cluster.  Whereas projected cluster-centric radii
set lower limits on the true distances of cluster members from the
cluster dynamical center, observed line-of-sight velocities say little
about their relative locations to the cluster, not to mention
obtaining a number of redshifts requires observationally expensive
spectroscopic surveys.  As the connections between the
morphology-radius and morphology-density relations and the
hierarchical picture of structure formation become clearer, it is of
great interest to better constrain the relations between the dynamical
states of clusters and their effect on constituent cluster galaxies.

\begin{figure}
\plotone{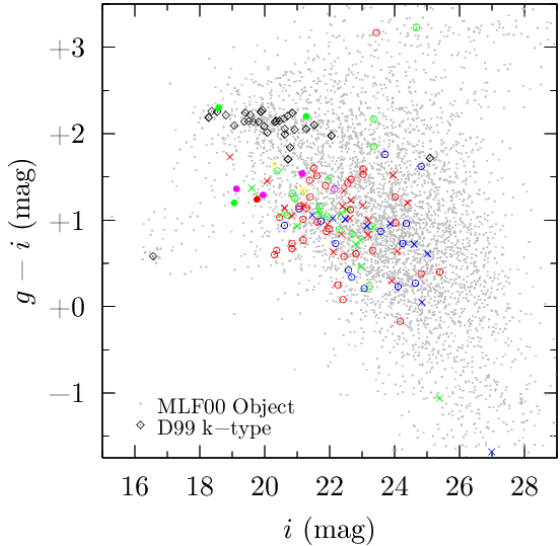}
\caption{ The (\gi)-$i$ color-magnitude diagram of all
\citetalias{mlf00} objects detected in their broadband images (gray
dots), \citetalias{dre99} k-type galaxies (open diamonds), and cluster
\fOII\ emitters (symbols as in \fig{fig.W0eVsLogSigma}).  The \gi\
colors and $i$ magnitudes of \citetalias{dre99} objects are determined
by those of spatially cross-correlated objects in the
\citetalias{mlf00} catalog, which includes the \gi\ colors of all the
objects detected in their broadband images.  In the color version, the
\citetalias{dre99} spectral types for the \fOII\ emitters
(\sect{sec.ClassificationBySpectralType}) from the best estimates of
the \fOII\ and \Hd\ equivalent widths are indicated by blue [e(b)],
green [e(c)], red [e(a)], yellow (k), and magenta (k+a/a+k).
[\emph{See the electronic edition of the paper for a color version of
this figure.}] }
\label{fig.GIClrMag}
\end{figure}

In this section, we analyze the kinematic states of cluster \fOII\
emitters.  This population is compared to the
spectroscopically-confirmed early-type galaxy population from the
\citetalias{dre99} survey to shed more light on the connections
between the dynamical states and the star-forming property of cluster
\fOII\ emitters.  Although the size of \citetalias{dre99}
spectroscopic sample is small, $N=31$ for k-type galaxies (see
\sect{sec.D99SpectralClassDefinitions} for definition of spectral
types), comparing the \fOII\ emitters to the galaxies composed of much
older stellar population observed in the same cluster illuminates the
kinematic segregation of galaxies with different ages often reported
in literature.  These two classes of galaxies are clearly separated in
the (\gi)-$i$ color-magnitude diagram (\fig{fig.GIClrMag}), and most
\citetalias{dre99} k-type galaxies belong to a well-defined cluster
red sequence.

\subsection{Velocity Distribution}
\label{sec.VelocityDistribution}

\begin{figure}
\plotone{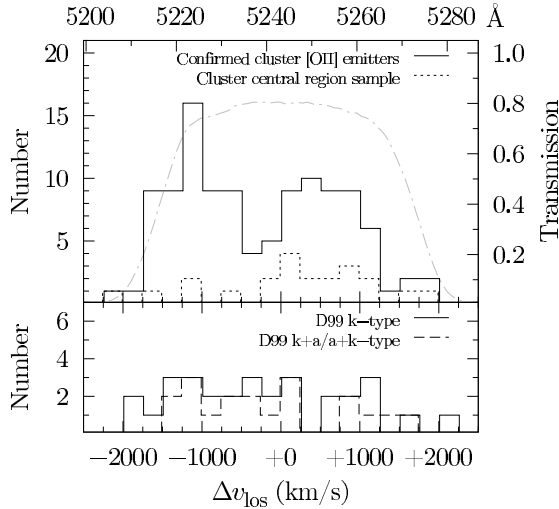}
\caption{ (\emph{Top}) Cluster rest-frame line-of-sight velocity
distribution of cluster \fOII\ emitters.  The solid histogram
corresponds to the spectroscopically-confirmed cluster members with
\fOII\ emission, and the short-dashed histogram is that of cluster
\fOII\ emitters within the common survey area with \citetalias{dre99}.
Top axis corresponds to the central wavelength of redshifted \fOII\
emission line.  The \citetalias{mlf00} on-band filter transmission
curve is shown in a gray dash-dotted line with the transmission
efficiency indicated on the right axis.  (\emph{Bottom}) The velocity
distribution of the confirmed Abell 851 members with a k- (solid
histogram) and k+a/a+k-type spectrum (long-dashed histogram) from
\citetalias{dre99}. }
\label{fig.VelocityHist}
\end{figure}

For each cluster \fOII\ emitter we computed its cluster-centric
line-of-sight velocity (or simply radial velocity) in the rest-frame
of cluster:
\[
\Delta\vlos \equiv c \frac{z - z_{\rm cl}}{1 + z_{\rm cl}} \ ,
\]
where $z$ is the observed redshift of the galaxy and $z_{\rm cl} =
0.4069$ for \object{Abell 851}.  \fig{fig.VelocityHist} shows the
velocity distribution of cluster \fOII\ emitters and
\citetalias{dre99} cluster member galaxies.  It is quite obvious that
the velocity distribution of our cluster \fOII\ emitters is not well
described by a Gaussian.  A virialized population is expected to have
a Gaussian velocity distribution around the cluster systemic velocity.
A K-S test on the distribution of \fOII\ emitters rejects a Gaussian
at a very high confidence level ($>99\%$).  A peculiar feature of the
velocity distribution of cluster \fOII\ emitters is the substantially
small fraction in low-velocity \fOII\ emitters (e.g.,
$-500~\km~\s^{-1} < \Delta\vlos < 0~\km~\s^{-1}$).  Whether
\citetalias{dre99} k-type galaxies show a similar deficit is not
clear; statistical tests involving the \citetalias{dre99} sample are
not very illuminating as its small sample size generally gives
inconclusive results.  For example, a two-sample K-S test of the
velocity distributions of all the \fOII\ emitters and
\citetalias{dre99} k-type galaxies with $-2500~\km~\s^{-1} <
\Delta\vlos < +2500~\km~\s^{-1}$ gives $\sim 92\%$ confidence that the
two distributions \emph{are} drawn from the same parent population.
Yet, if we use the subsamples restricted spatially within the common
survey area of $-3\farcm74 \le \Delta{\rm RA} \le +3\farcm97$ and
$-2\farcm42 \le \Delta{\rm DEC} \le +1\farcm91$,\footnote{The
coordinates are given with respect to \clustercenter.} the same
statistical test rejects at the $97\%$ confidence level the hypothesis
that they are drawn from the same parent distribution.  At least
\citetalias{dre99} k-type galaxies do not show as pronounced a
depletion near the cluster systemic velocity and might represent a
virialized population; this is not a baseless assumption, since the
\citetalias{dre99} k-type galaxies appear to form a well-defined
cluster red-sequence (\fig{fig.GIClrMag}), and such galaxies are
likely to trace gravitationally-bound substructures (although it is
almost certainly the case that \object{Abell 851} has not been
virialized over the entire cluster scale).

The biweight estimator and the jackknife method are used to estimate
the velocity dispersions $\sigma_{v}$ and their uncertainties
\citep{bee90}: $1160 \pm 140~\km~\s^{-1}$ and $1010 \pm
50~\km~\s^{-1}$ for \citetalias{dre99} k-type galaxies and cluster
\fOII\ emitters, respectively.  We must note, however, that
\fig{fig.VelocityHist} shows that a significant number of cluster
\fOII\ emitters with $\Delta\vlos \la -1000~\km~\s^{-1}$ escaped
detection by the \citetalias{mlf00} narrowband search due to the
sharply dropping transmission efficiency.  This would lead to a severe
underestimate of the true velocity dispersion for the cluster \fOII\
emitting population; their velocity dispersion should therefore be
taken as a lower limit.

A simple kinematical treatment of infalling and virialized cluster
galaxies in a cluster-scale potential leads to $\left|T/V\right|
\approx 1$ and $\left|T/V\right| \approx 1/2$, respectively, where $T$
and $V$ are the kinetic and potential energies of galaxies.  In this
picture, the velocity dispersions of infalling and virialized galaxies
are expected to follow the relation $\sigma_{v,{\rm infall}} \approx
2^{1/2}\sigma_{v,{\rm vir}}$.  Having a higher velocity dispersion
compared to that of virialized cluster members therefore is often
invoked as a basic attribute of an infalling galaxy population.
Previous studies generally found such kinematic segregations of
cluster galaxies, with star-forming galaxy populations having higher
velocity dispersions \citep[e.g.,][]{col96,dre99,dia01}, although
\citet{rin05} found little evidence for the generality of this trend
using their sample of emission-dominated and absorption-dominated
galaxies in virialized regions of clusters.  Such disagreements only
illuminate the intrinsic difficulty of minimizing the effects of
biases in cluster-scale surveys of galaxies, which require a
substantial amount of spectroscopy.

In short, the cluster \fOII\ emitters appear not to be virialized,
since their velocity distribution does not follow a Gaussian around
the cluster systemic velocity.  Although we cannot rigorously conclude
due to the small sample size of \citetalias{dre99} galaxies, their
velocity distributions restricted to the common survey area show some
evidence of kinematic segregation between the cluster \fOII\ emitters
and the \citetalias{dre99} k-type galaxies.  The non-virialized state
of \fOII\ emitters naturally arises when they are at the infalling
phase and have not spent a long time within the cluster gravitational
potential.

\subsection{Spatial Distribution}
\label{sec.SpatialDistribution}

\begin{figure*}
\plotone{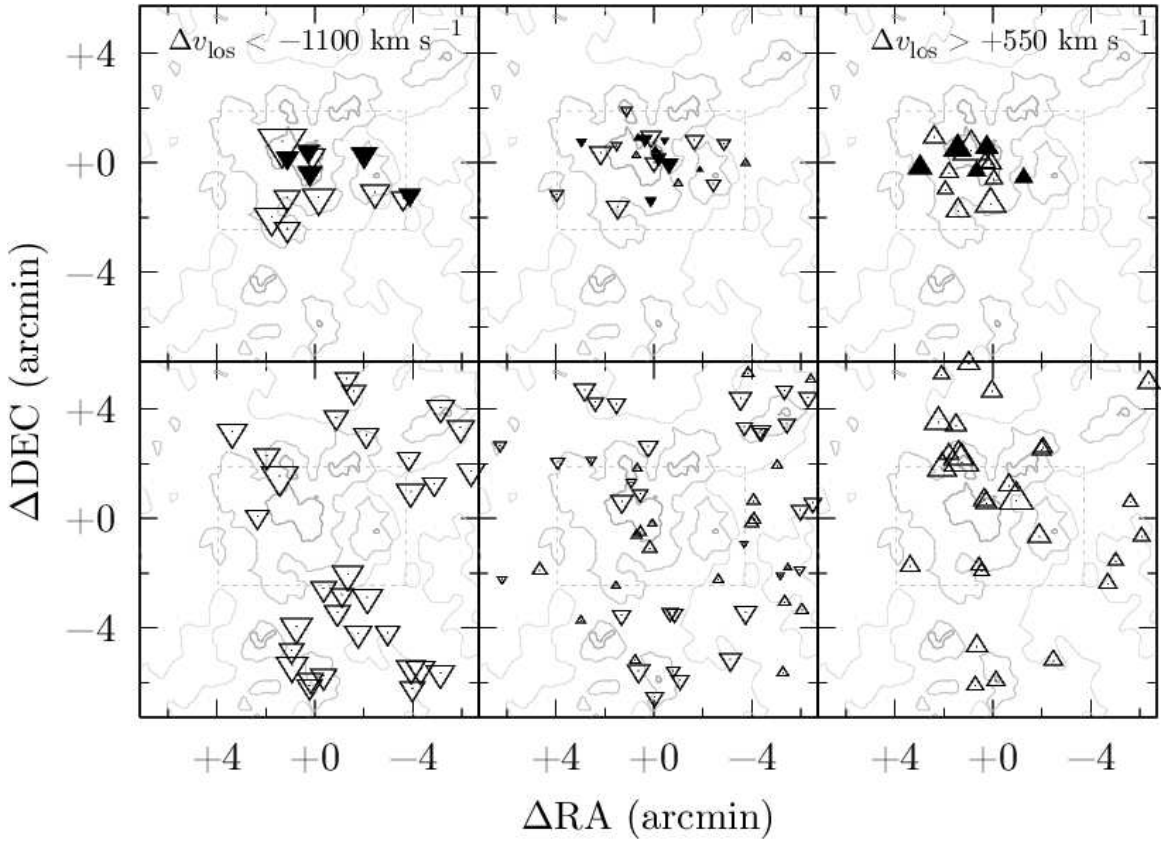}
\caption{ Spatial distributions of cluster \fOII\ emitters
(\emph{bottom row}) and \citetalias{dre99} galaxies (\emph{top row})
with k- (open triangles) and k+a/a+k-type (filled triangles) spectrum.
Lack of \citetalias{dre99} galaxies beyond the area enclosed by a gray
dashed box is due to the smaller survey area of \citetalias{dre99}.
The gray contours indicate the local galaxy densities at $5$, $50$,
and $150~\Mpc^{-2}$ computed from \citetalias{kod01} cluster members.
In the columns the galaxies are binned by their velocity: $\Delta\vlos
< -1100~\km~\s^{-1}$ (\emph{left}), $-1100~\km~\s^{-1} < \Delta\vlos <
+550~\km~\s^{-1}$ (\emph{middle}), and $\Delta\vlos >
+550~\km~\s^{-1}$ (\emph{right}).  Upright and inverted triangles
denote redshift ($\Delta\vlos > 0~\km~\s^{-1}$) and blueshift
($\Delta\vlos < 0~\km~\s^{-1}$) with respect to the cluster systemic
velocity, respectively.  The symbol size is roughly proportional to
$\left|\Delta\vlos\right|$.  }
\label{fig.VelocitySpatialDist}
\end{figure*}

\fig{fig.VelocitySpatialDist}, on the other hand, shows clear evidence
for a kinematic segregation in terms of spatial distributions of the
cluster \fOII\ emitters and \citetalias{dre99} k-type galaxies.  In
contrast to \citetalias{dre99} k-type galaxies, \fOII\ emitters are
only sparsely distributed in the dense, cluster core region; there are
very few \fOII\ emitters inside or near the contour of the highest
local galaxy density.  This trend holds true in all the velocity bins
in the figure.  In \fig{fig.EmissFracVsLogSigma} we have shown that
the \fOII\ emitter fraction is a strong function of local galaxy
density, and \fig{fig.VelocitySpatialDist} only visualizes the
well-established empirical relation.  In wide-field imaging of
\object{Abell 851}, the red-sequence galaxies are shown to trace
similar large-scale structures and substructures as observed via the
photometric redshift technique \citep{kod05}, so it is of no surprise
that the \citetalias{dre99} k-type galaxies are found abundantly in
dense regions.

The very skewed distribution of blueshifted ($\Delta\vlos <
0~\km~\s^{-1}$) \fOII\ emitters is also evident in the southern
($\Delta{\rm RA} \simeq +0\arcmin$ and $\Delta{\rm DEC} \simeq
-6\arcmin$) and north-western regions ($\Delta{\rm RA} \simeq
-4\arcmin$ and $\Delta{\rm DEC} \simeq +4\arcmin$).  Systematically
redshifted \fOII\ emitters are present just in the north of the
cluster central region as well.  \object{Abell 851} is known for the
supposed filamentary structures extending out from its central region
in these directions \citepalias{kod01}.  The skewed distribution of
$\Delta\vlos$ indicate the \fOII\ emitters are part of infalling
substructures, especially when spatially clustered as seen clearly in
the figure.

In short, we find that the cluster \fOII\ emitters tend to avoid the
densest environments, but their skewed velocity and spatial
distributions strongly suggest they are moving systematically with
respect to the cluster dynamical center.  Some of the spatially
non-clustered \fOII\ emitters may be field interlopers.

\subsection{Infalling or Backsplash Population?}
\label{sec.InfallingOrBacksplashPopulation}

True positions of the galaxies relative to the cluster cannot be
determined from their redshift alone, however, and in addition to the
possibility of seeing field interlopers, even cluster members beyond
the virial radius ($R_{\rm vir}$) could also have gone through the
cluster central region once or more in the past, constituting a
so-called backsplash population \citep{bal00, mam04}.  According to
the numerical simulations, about half of the galaxies within $1 <
R/R_{\rm vir} < 2$ can belong to a backsplash population by the
present epoch, $z=0$.  The backsplash scenario has been invoked to
explain the presence of \HI\ deficient galaxies beyond virial radii of
nearby galaxy clusters, where the galaxies appear to be ram-stripped
of their \HI\ supplies during the passage through dense intracluster
gas of cluster cores \citep{sol02, san02}.  Since the inferred
physical mechanisms that transform cluster galaxies vary for infalling
and backsplash populations, finding their dynamical state helps
constraining SFHs of cluster galaxies.

\begin{figure}
\plotone{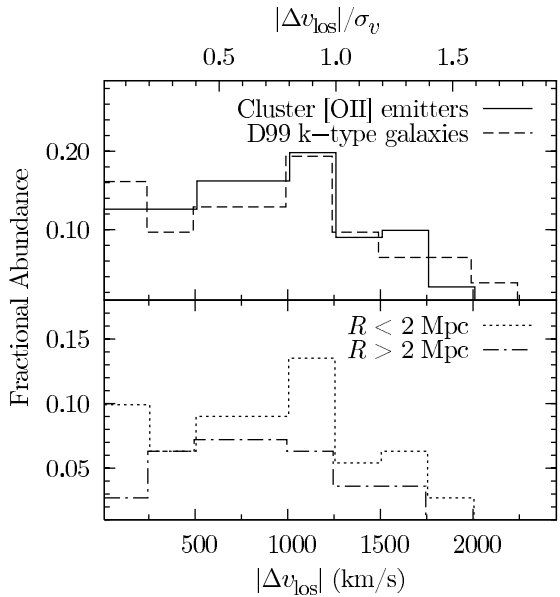}
\caption{ (\emph{Top}) Distribution of $\left|\Delta\vlos\right|$ for
cluster \fOII\ emitters (solid histogram) and \citetalias{dre99}
k-type galaxies (dashed histogram).  The top axis shows the velocity
normalized to an estimate of cluster velocity dispersion $\sigma_v =
1260~\km~\s^{-1}$.  The fractional abundance is normalized to the
total number of galaxies within each sample to facilitate comparison.
Lack of low-$\left|\Delta\vlos\right|$ galaxies is a signature of
infalling population \citep[cf., Fig.~8 of][]{gil05}.  We note that
the samples are not restricted to $1 < R/R_{\rm vir} < 2$.
(\emph{Bottom}) The velocity distributions of the cluster \fOII\
emitters inside (short-dashed) and outside (dash-dotted) the fiducial
virial radius of $2~\Mpc$.  The center of the cluster was taken to be
at \clustercenterXray, the brightest point in X-ray surface brightness
from \citet{def03}. }
\label{fig.VelocityHistAbs}
\end{figure}

The simulations carried out by \citet{gil05} have shown that
backsplash and infalling populations exhibit kinematically distinct
velocity distributions beyond virial radii, where a backsplash
population has about a factor of two smaller cluster-centric
velocities than infalling satellites.  They have also shown that the
distribution of cluster-centric line-of-sight velocities like
\fig{fig.VelocityHist} could imply the presence or absence of a
backsplash population.  Along this line, we show
\fig{fig.VelocityHistAbs} to facilitate a comparison to Fig.~8 of
\citet{gil05}.  The depletion of low-$\left|\Delta\vlos\right|$ \fOII\
emitters shows quite a remarkable similarity to the distribution
expected of an infalling population in their simulation.  It should
also be noted that unlike Fig.~8 of \citet{gil05} the sample used in
the top figure is not restricted to those within the projected radii
of $1 < R/R_{\rm vir} < 2$, where the difference would be more
pronounced between backsplash and infalling populations.  In the same
figure, the \citetalias{dre99} k-type galaxies also appear to show a
peak at a higher velocity, the signature distribution of infalling
population, just as the cluster \fOII\ emitters do.  However, we
believe there are important differences, since the \citetalias{dre99}
k-type galaxies are drawn from a much smaller field of view
(\fig{fig.VelocitySpatialDist}), are significantly brighter
(\fig{fig.GIClrMag}), and are not necessarily representative in a
statistical sense without applying proper corrections \citep{pog99}.
We also lack \citetalias{dre99} galaxies at large projected cluster
radii.  Furthermore, the peak around $\left|\Delta\vlos\right| \simeq
1000~\km~\s^{-1}$ do not appear to indicate a real over-abundance of
\emph{general} galaxy population around that velocity, since they are
not spatially coincident; see \fig{fig.VelocityHist} for the cluster
central region sample.  For these reasons, we believe the depletion of
low-$\left|\Delta\vlos\right|$ galaxies is robust for the cluster
\fOII\ emitters, yet the small sample size limits our ability to
interpret the distribution of \citetalias{dre99} k-type galaxies.

In the bottom of \fig{fig.VelocityHistAbs}, we show the velocity
distributions of \fOII\ emitters inside and outside the fiducial
cluster virial radius of $2~\Mpc$, the latter enabling a more direct
comparison to the \citeauthor{gil05} simulations.  We caution that the
distributions shown here are rather sensitive to the choice of the
fiducial virial radius, since the projected distribution of the
cluster \fOII\ emitters are highly inhomogeneous as discussed earlier.
Recently both \citet{rin05} and \citet{pim06} carried out similar
comparisons using composite galaxy samples from nearby clusters,
finding galaxies in infall regions are a mixture of infalling and
backsplash populations.  Since the \citeauthor{gil05} results are for
massive, moderately relaxed clusters at $z\sim 0$, these works
produced more direct comparisons.  In contrast to these authors'
findings, the ``peaks'' of the velocity distributions of our cluster
\fOII\ emitters extend toward higher velocities up to
$|\Delta\vlos|/\sigma_v \simeq 1$, whereas the galaxies in both
\citeauthor{rin05} and \citeauthor{pim06} show a steep decline in
numbers at $|\Delta\vlos|/\sigma_v > 0.5$.  For the $R>2~\Mpc$ sample,
we also observe a more pronounced depletion of low-\vlos\ \fOII\
emitters.  These trends imply that the total \fOII-emitting population
is probably a mixture of infalling and backsplash populations, yet the
velocity distributions are generally broadened toward higher
velocities than observed in local clusters, favoring a higher
abundance of infalling objects.

In short, although the survey area of our spectroscopic sample is not
as large as some of the most recent wide-field studies of galaxy
clusters, e.g., \citet{tre03} and \citet{mor05} go out to radii of
$\sim 5~\Mpc$ in \object{CL0024+1654}, the analysis of their
kinematics suggests that a significant fraction of the cluster \fOII\
emitters are likely to be still in the infalling phase.

\section{Classification by Spectral Type}
\label{sec.ClassificationBySpectralType}

To probe transformation of galaxy properties, we need observable
indicators that are sensitive to the age of underlying stellar
population.  Classifying galaxies in terms of \fOII-\Hd\ equivalent
width plane is one way to put galaxies in the context of evolutionary
sequence and has been used in literature \citep[e.g.,][]{dre99,bal99}.
The equivalent widths of \fOII\ emission and \Hd\ absorption,
respectively, are sensitive to the on-going star-forming activity and
the presence of intermediate-age stars, i.e., A- and early F-type star
populations which live up to $\sim 1.5~\Gyr$ with characteristically
strong Balmer absorption lines.  In the conventional interpretation of
post-starburst galaxy spectra, the presence of strong Balmer
absorption lines signals the recent cessation within the past $\sim
1.5~\Gyr$ of star-forming activity after which the intermediate-age
stellar population makes a dominant contribution to the integrated
optical/UV galaxy spectrum.  Therefore, the \fOII-\Hd\ plane and its
variants have been useful for classifying a galaxy spectrum in
relation to some mechanisms which halt or significantly reduce the
amount of star-forming activity within a galaxy in its recent past.

\subsection{\citetalias{dre99} Spectral Class Definitions}
\label{sec.D99SpectralClassDefinitions}

\begin{deluxetable}{ccccc}
\tabletypesize{\footnotesize}
\tablecaption{\citetalias{dre99} Spectral Classification}
\tablewidth{0pt}
\tablehead{
\colhead{} &
\multicolumn{2}{c}{Definition} &
\multicolumn{2}{c}{Cluster \fOII\ Emitters} \\
\cline{2-3}\cline{4-5} \\
\colhead{} &
\colhead{$W_0(\fOII)$} &
\colhead{$W_0(\Hd)$} &
\colhead{$N(\snr>2)$\tablenotemark{a}} &
\colhead{$N({\rm all})$\tablenotemark{b}} \\
\colhead{Spectral Class} &
\colhead{($\ang$)} &
\colhead{($\ang$)} &
\colhead{} &
\colhead{}
}
\startdata
e(b)    & $<-40$        & $<+4$ &  3 & 23 \\
e(c)    & $-40$ -- $-5$ & $<+4$ & 14 & 29 \\
e(a)    & $<-5$         & $>+4$ & 40 & 53 \\
k+a/a+k & $>-5$         & $>+3$ &  2 &  4 \\
k       & $>-5$         & $<+3$ &  1 &  2
\enddata

~\tablecomments{ The spectral indices $W_0(\fOII)$ and $W_0(\Hd)$ are
measured with the flux-summing method \citepalias{sat06}. }

\tablenotetext{a}{ The number of cluster \fOII\ emitters for which the
classification is done with the equivalent width measurements at $\snr
> 2$. }

\tablenotetext{b}{ The number of cluster \fOII\ emitters with each
classification from the full sample. }

\label{tab.D99SpecClassScheme}
\end{deluxetable}

We adopt the \citetalias{dre99} spectral classification scheme to
study the SFHs of cluster \fOII\ emitters
(\tab{tab.D99SpecClassScheme}).\footnote{ For a few objects the
$\fOII\lam3727$ or \Hd\ line moved out of observed spectral range, in
which case their spectra were classified into ``N/A''-type. }  We
stress that by convention the measured \Hd\ equivalent width in the
\fOII-\Hd\ plane is the sum of emission and absorption equivalent
widths.  This means that $W_0(\Hd) < 0~\ang$ does not mean the line is
totally seen in emission; it merely means the strength of emission
equivalent width is greater than that of absorption at \Hd.  Although
we could separate the emission and absorption components at \Hd\ in
many cases, we followed the conventional index definition to
facilitate comparisons to literature.  We used the equivalent widths
measured using the flux-summing method for spectral indices
\citepalias{sat06}.  We now briefly review the conventional
interpretation of each spectral class.  For more detail, readers are
referred to the original papers by \citetalias{dre99} and
\citet{pog99}.

\emph{e(b)-type}.---Galaxies in this class have a strong \fOII\
emission, indicative of vigorous on-going star-forming activity.
Low-lying recombination lines of the Balmer series may also be seen in
emission.  The duration of a starburst cannot be longer than about a
few $100~\Myr$, however, since after that timescale the buildup of
continuum flux near the line decreases the \fOII\ emission equivalent
width to a degree that the spectrum would be classified into
e(c)-type.

\emph{e(c)-type}.---Typical e(c) galaxies are a star-forming galaxy
with only a moderate star-forming activity at a relatively constant
rate.  Local late-type spirals typically fit in this category.

\emph{e(a)-type}.---Typical galaxies in this class have strong Balmer
absorption lines as well as moderate to substantial $\fOII\lam3727$
emission.  Based on the comparison with a local sample and high degree
of dust extinction implied by low $\fOII/\Ha$ line ratios,
\citet{pog00} have associated typical e(a)-type galaxies to dusty
starburst galaxies in which the light from the youngest starbursting
population is much more extincted than that from the intermediate to
old age stellar populations \citep[see also][]{pog01}.

\emph{k+a/a+k-type}.---So-called post-starburst galaxies (also called
E+A or \Hd-strong galaxies) with little or no \fOII\ emission and
strong Balmer absorption lines belong to this class.  This class of
galaxies is often studied as a missing link between passively-evolving
galaxies in clusters and actively--star-forming galaxies abundantly
observed outside clusters.

\emph{k-type}.---Passively evolving, non--star-forming galaxies like
local ellipticals belong to this class.  A typical k-type spectrum is
dominated by the light from K-type star population, hence the
designation.

\begin{figure}
\plotone{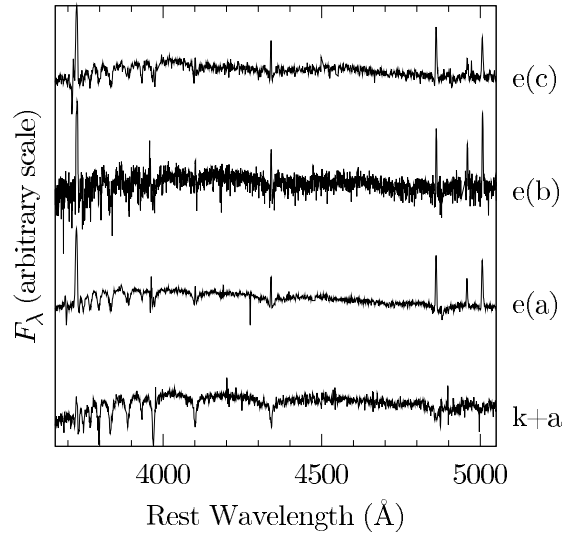}
\caption{ A set of observed high-\snr\ spectra classified into each
\citetalias{dre99} spectral class.  The regions affected by very poor
sky subtractions are masked from each spectrum in this figure.  Note
that a bright skyline and telluric absorption lines generally
contaminated redshifted \Hb\ absorption continua, substantially
increasing the uncertainties in the \Hb\ line flux measurements. }
\label{fig.TypicalSpectra}
\end{figure}

A few representative spectra, actually observed in our spectroscopic
survey, are shown in \fig{fig.TypicalSpectra} with their corresponding
\citetalias{dre99} spectral class.  \tab{tab.D99SpecClassScheme} also
lists the numbers of cluster \fOII\ emitters in each spectral class.
As expected, most \fOII\ emitters belong to one of the emission-line
classes [i.e., e(b), e(a), and e(c)], but six objects, all of which do
show weak \fOII\ emission, are classified into non--emission-line
classes.  By making a cut at $W_0(\fOII) > -5~\ang$, the
\citetalias{dre99} classification scheme does not strictly remove the
contamination of emission-line objects into their ``passive'' classes.
Among the four k+a/a+k type objects, three are strong
Balmer-absorption systems with residual \fOII\ emission.  One of the
two k-type objects appears to have an early-type galaxy spectrum but
also has \fOII\ and Balmer emission lines.  The remaining two objects
appear genuine ambiguous classifications due to poor-\snr\ spectra.

\subsection{Distribution of \fOII\ Emitters by Spectral Class}
\label{sec.DistributionOfOIIEmittersBySpectralClass}

\begin{figure*}
\plotone{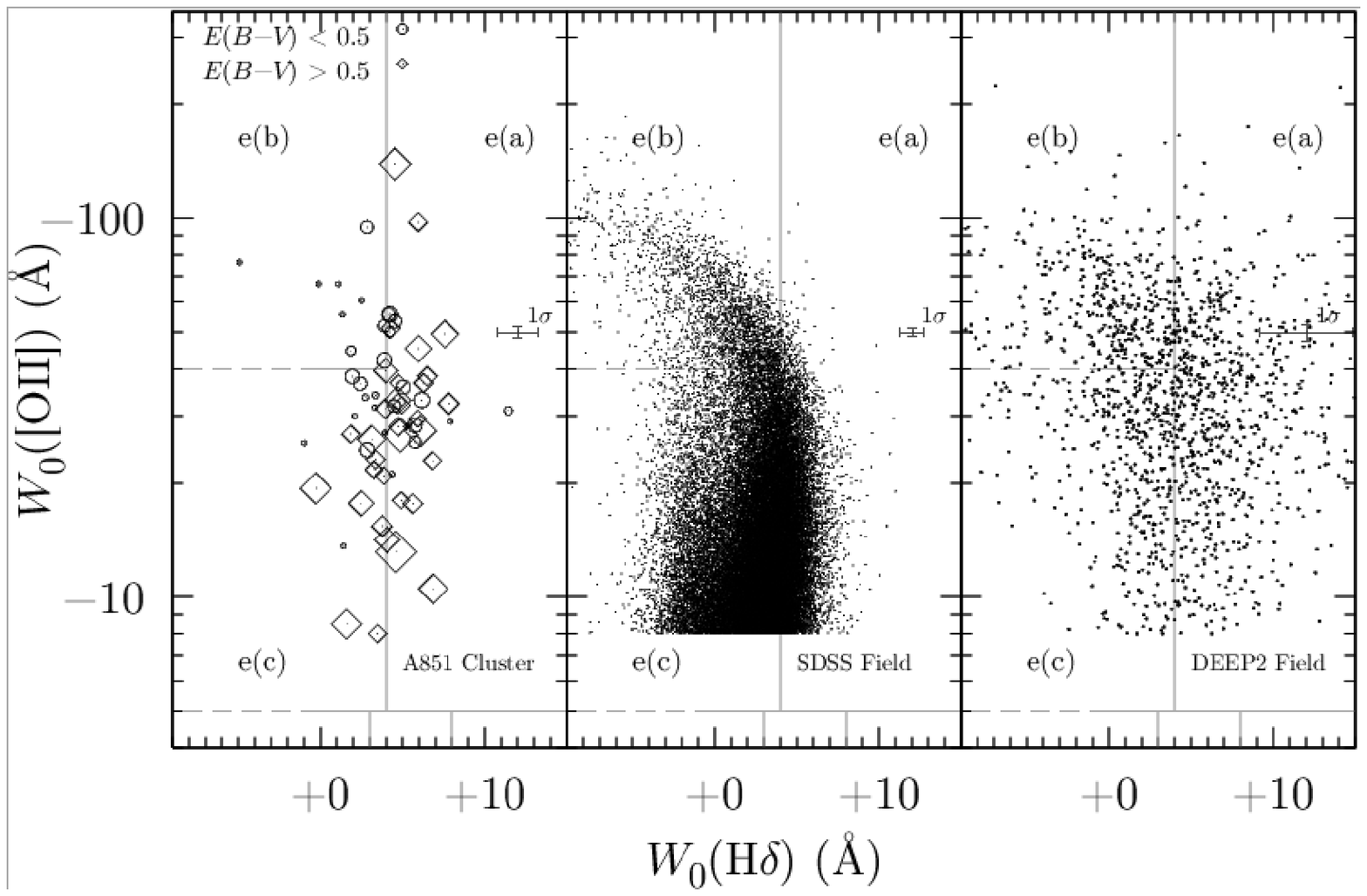}
\caption{ The distributions of \fOII\ and \Hd\ equivalent width for
$W_0(\fOII) < -8~\ang$ objects for magnitude-limited samples.  See
\sect{sec.DistributionOfOIIEmittersBySpectralClass} for the selection
methods for \SDSS\ and DEEP2 field samples.  For the Abell 851 \fOII\
emitters, their reddening \EBV\ is coded by circles [$\EBV < 0.5$] and
squares [$\EBV > 0.5$], and their size is proportional to \EBV.  The
$1\sigma$ error bars indicate the median value for a measurement
within each sample. }
\label{fig.SCO2VsHd}
\end{figure*}

\begin{deluxetable*}{lcccc}
\tabletypesize{\footnotesize}
\tablecaption{Comparison Samples of Field \fOII\ Emitters}
\tablewidth{0pt}
\tablehead{
\colhead{Survey} &
\colhead{$z$ Range} &
\colhead{Magnitude Cut} &
\colhead{Luminosity Function} &
\colhead{Number of Spectra}
}
\startdata
\SDSS\ (DR4)\tablenotemark{a}  & $0.09$--$0.11$ & $i < 23$\tablenotemark{b} &
\citet{ball05}\tablenotemark{c} & 41613 \\
\CFRS\tablenotemark{d}         &  $0.2$ -- $0.7$  & $I({\rm AB}) < 22.5$ &
N/A\tablenotemark{e}            &    51 \\
DEEP2 (DR1)\tablenotemark{f}   &  $0.8$ -- $1.0$  & $R({\rm AB}) < 24.1$ &
\citet{wil05}\tablenotemark{g}  &  1492
\enddata

\tablecomments{ The spectral indices $W_0(\fOII)$ and $W_0(\Hd)$ are
measured with the flux-summing method \citepalias{sat06}.  Each field
sample consists of galaxies with $W_0(\fOII) < -8~\ang$.  }

\tablenotetext{a}{ Only extended sources which are spectroscopically
classified as galaxy with ${\rm zconf} > 0.85$ are included. }

\tablenotetext{b}{ The magnitude limit of \SDSS\ survey is
$i\approx21.3$; we include fainter galaxies to increase the number
of galaxies near $M_B \approx -19$.  All photometry in our \SDSS\
samples are Petrosian magnitudes. }

\tablenotetext{c}{ eClass PCA bivariate luminosity function (their
Fig.~5).  Within each magnitude bin and eClass range, an additional
correction for the fraction of galaxies with $W_0(\fOII) < -8~\ang$ is
applied to remove the contribution to the luminosity function from
non--emission-line galaxies. }

\tablenotetext{d}{ Only objects with their confidence class $0$--$4$
\citep{lef95} are included.  As the \CFRS\ spectra were obtained at
considerably different spectral resolutions ($40~\ang$) than other
surveys ($\sim 3~\ang$ for \SDSS\ and $\sim 1~\ang$ for DEEP2), we
used slightly different windows of $3653~\ang$--$3706~\ang$,
$3706~\ang$--$3748~\ang$, and $3748~\ang$--$3801~\ang$ (\fOII) and
$4030~\ang$--$4078~\ang$, $4078~\ang$--$4126~\ang$, and
$4126~\ang$--$4170~\ang$ (\Hd) for the blue continuum, line, and red
continuum, respectively.}

\tablenotetext{e}{ No attempts have been made to correct the sample
using a luminosity function. }

\tablenotetext{f}{ Only objects with ${\rm zquality} \ge 3$ are
included. }

\tablenotetext{g}{ Schechter function for blue galaxies (their
Table~4). }

\label{tab.FieldSamples}
\end{deluxetable*}

For our sample of cluster \fOII\ emitters, the limiting rest-frame
\fOII\ equivalent width is $\simeq -8~\ang$ at $i \simeq 23$, so there
are few galaxies of k- or k+a/a+k-type, which must have $W_0(\fOII) >
-5~\ang$ by definition; most \fOII-selected galaxies belong to one of
the emission-line galaxy types: e(b)-, e(c)-, and e(a)-types.  In
\fig{fig.SCO2VsHd} we show the distribution in the \fOII-\Hd\ plane
for \object{Abell 851} cluster \fOII\ emitters with $W_0(\fOII) <
-8~\ang$ and $i < 23$.  To see how the distribution compares to
``field'' \fOII\ emitters, we measured the same spectral indices using
the fourth data release of the Sloan Digital Sky Survey
\citep[\SDSS;][]{ade06}, the Canada-France Redshift Survey
\citep[\CFRS;][]{lil95, ham97}, and data release one of the DEEP2
Galaxy Redshift Survey.\footnote{http://deep.berkeley.edu/} For each
survey, we only included galaxies with $W_0(\fOII) < -8~\ang$ to
compute the relative abundances of spectral types.
\tab{tab.FieldSamples} summarizes the basic characteristics of the
samples.  The typical size of the cluster \fOII\ emitters is
$2''$--$3''$ ($\simeq 13~\kpc$ at $z=0.4$), observed with $1''$-width
slit in our spectroscopic program.  The spectroscopic apertures in
\SDSS\ ($3''$ diameter fiber), \CFRS\ ($1\farcs75$-width slit), and
DEEP2 ($1''$-width slit) sample the regions of galaxies roughly within
$5$--$6~\kpc$, $>11~\kpc$, and $>8~\kpc$, respectively, at the median
redshift of each survey.  Although we do not make an attempt to
correct for the differences, the regions sampled by the surveys are
comparable or at least greater than our spectroscopic aperture.  All
the surveys are magnitude-limited to different depths, so field
samples can only be constructed to a depth brighter than the
completeness limit of the cluster \fOII-emitting sample.

We constructed rough volume-limited samples by scaling the number of
galaxies within similar rest-frame $B$-band absolute magnitudes by
suitable luminosity functions from existing literature.  We used
software package kcorrect \citep[v4\_1\_4;][]{bla03} to obtain
absolute magnitudes from a set of available broadband photometry
(corrected only for Galactic extinction) for each galaxy survey.  The
cluster \fOII\ emitters near the \citetalias{mlf00} narrowband survey
limit have $M_B({\rm Vega}) \simeq -17.5$ (before correcting for
internal extinction).  Since the completeness severely decreases
toward that magnitude, we limited our comparisons roughly to the
survey limit of our DEEP2 sample, $M_B < -19$; for both the
\SDSS\ and DEEP2 samples, the rough completeness limit in terms of
absolute magnitude is $M_B \sim -21$, so the uncertainties in galaxy
fractions increase significantly toward the faint end of luminosity
function for that reason.  We caution that the above procedure was
not followed for the \CFRS\ sample because of limited availability of
photometric data and significantly lower-resolution spectra; thus the
\CFRS\ data points should only be used for a rough comparison.  We
further note that using purely magnitude-limited samples did not
significantly change the overall trends to be discussed now.

Looking at the \SDSS\ field sample in the \fOII-\Hd\ plane is
illuminating because of its large sample size and relatively small
measurement uncertainties.  We notice the distribution is dominated by
typical star-forming galaxies [e(c)], with relatively small pure
starbursting population [e(b)].  The galaxies with e(a)-type spectrum
appear to form some extension, toward stronger \Hd\ absorptions, of
general star-forming galaxy population with e(c)-type spectrum.  The
DEEP2 field sample generally suffers from large uncertainties; the
distribution of galaxies with low \fOII\ equivalent width does not
appear to extend well down below $\sim -10~\ang$ in contrast to the
\SDSS\ sample, which indicates more vigorous star-forming activity
expected of galaxies at high redshift.

\begin{figure}
\plotone{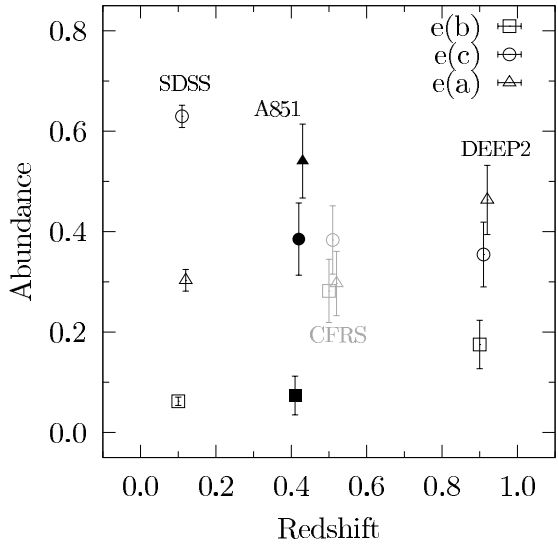}
\caption{ Fractional abundance of each \citetalias{dre99} spectral
class relative to all $W_0(\fOII) < -8~\ang$ galaxies in each survey.
Filled symbols denote the Abell 851 \fOII\ emitters.  Each sample is
shown roughly at its median redshift with a slight offset for
clarity. }
\label{fig.FracSCVsZ}
\end{figure}

In \fig{fig.FracSCVsZ} we show the relative abundance of each
\citetalias{dre99} class within each sample of $W_0(\fOII) < -8~\ang$
galaxies at different redshift.  As expected, the e(b)-type galaxy
fraction was higher at high redshift, reflecting the fact that
triggering of young starbursts was more frequent as the universe had a
greater amount of star-forming activity out to $z\sim2$.  A similar
redshift trend is seen in the e(a)-type fractions in the field galaxy
samples, but in \object{Abell 851} the e(a) fraction is higher than or
at least comparable to that in the field within the probed redshift
range.  It is important to note that the plot shows the fractional
abundance of different spectral classes within each $W_0(\fOII) <
-8~\ang$ galaxy samples.  Since the \fOII\ emitters constitute a
progressively smaller fraction of galaxies toward lower redshift
\citep{nak05}, the observed trend of e(b) and e(a) galaxies in the
field is stronger, if the fractions are computed within the full
galaxy samples.  For the same reason, the e(c) fractions within the
full samples do not evolve very much in the field, if we adopt the
\fOII\ galaxy fraction in \citeauthor{nak05}\ As noted earlier,
e(a)-type spectrum has been attributed to dusty starburst galaxies.
If this identification is true, even the e(a) galaxies constitute a
starbursting population, implying that starburst phenomena in
\object{Abell 851} are at least as frequent as $z\sim1$ field \fOII\
emitters.

In computing fractional abundance of each emission-line galaxy class,
we opted to take each data point in the \fOII-\Hd\ plane as a
two-dimensional Gaussian probability distribution with its best
estimates and standard deviations given by the measurements of \fOII\
and \Hd\ equivalent widths and their uncertainties.  Each data point
integrates to unity over the entire \fOII-\Hd\ plane, but low-quality
measurements give rise to some probability for each point to leak into
other spectral classes than that designated solely by their best
estimates.  A significant number of both \object{Abell 851} and field
\fOII\ emitters have spectra with low \snr---about a few per pixel for
\object{Abell 851} \citepalias{sat06} and even smaller for the DEEP2
sample; by following the above procedure, we avoid giving too much
confidence to the classification using the best estimates of such
low-\snr\ measurements.  Dividing the sum of integrated probabilities
in each spectral class by the total number of galaxies in the
\fOII-\Hd\ plane yields the fractional abundance.  For \object{Abell
851} sample, we have taken into account the spectroscopic target
selection bias by multiplying each \gi\ color subsample by an
appropriate correction factor \citepalias{sat06}.

\section{Star Formation History of Cluster \fOII\ Emitters}
\label{sec.SFHOfClusterOIIEmitters}

Analyzing the cluster \fOII\ emitters in terms of possible
evolutionary scenarios requires modeling their spectral evolution
using stellar population synthesis models.  The \citetalias{dre99}
spectral classification scheme employs a combination of stellar and
nebular lines that are sensitive to the time evolution of galaxy
spectra over a wide range in stellar age.  Hence our spectral modeling
considers the time evolution of synthetic spectra of underlying
stellar populations as well as corresponding nebular emission.
Furthermore, combining these two components enabled us to incorporate
the effect of differential dust extinction on stellar and nebular
components, which appears essential for reproducing e(a) spectrum
\citep{pog01}.  The detail of our modeling scheme is outlined in
\append{sec.SpectralEvolutionModeling}.

\subsection{Time Evolution in the \fOII-\Hd\ Plane}
\label{sec.TimeEvolutionInTheOII-HdPlane}

\begin{figure}
\plotone{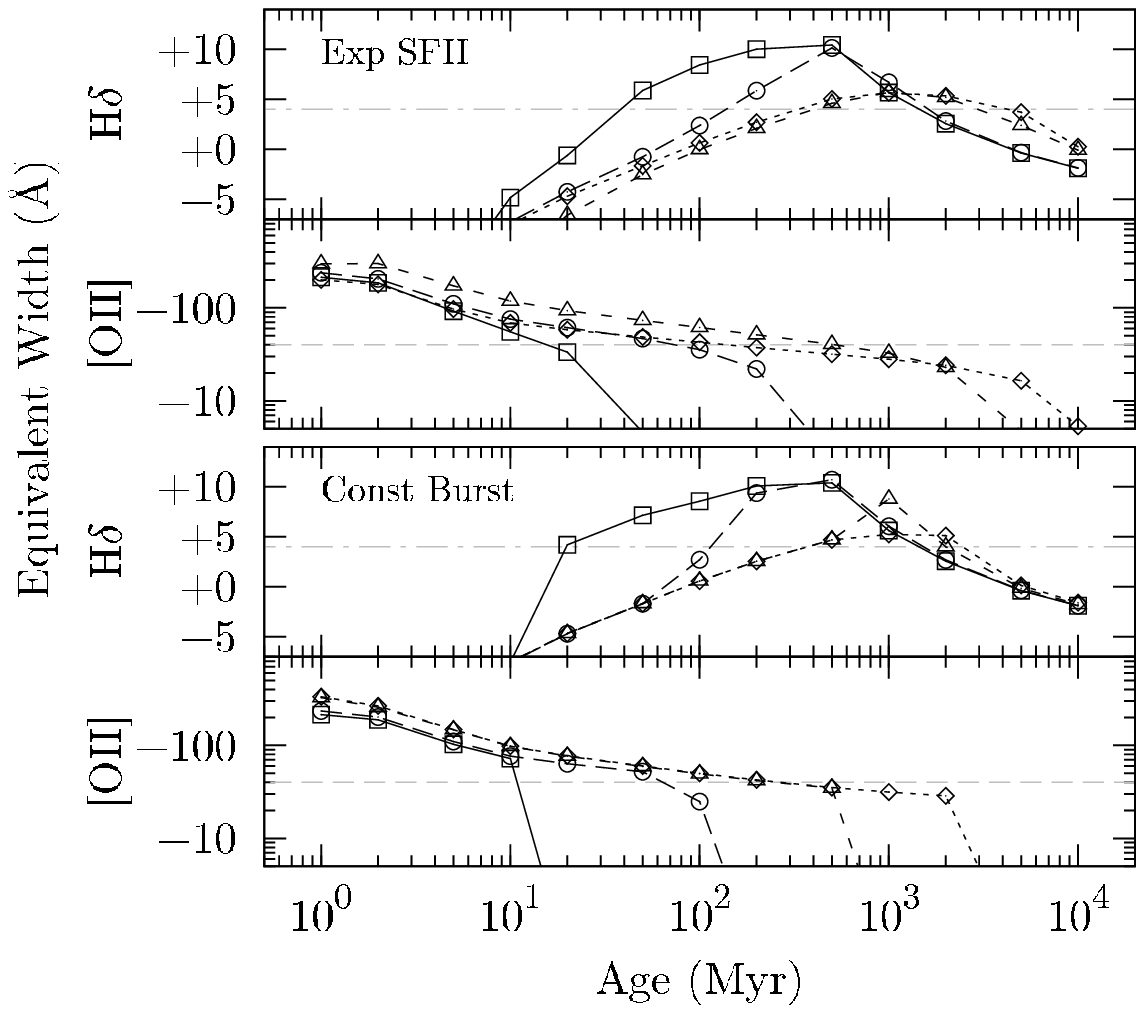}
\caption{ The \fOII\ and \Hd\ equivalent widths as a function of age
for constant burst (\emph{bottom}) and exponentially-declining SFH
(\emph{top}) models with the fiducial parameters $\EBV = 0.5$ and
$\EBVr = 0.44$ (\append{sec.SpectralEvolutionModeling}).  The models
were run for star-forming timescales of $\tau = 10~\Myr$ (solid line
with square symbols), $100~\Myr$ (dashed with circles), $1~\Gyr$
(short-dashed with triangles), and $2~\Gyr$ (dotted with diamonds)
models.  The measurements are made at the ages ($1$, $2$, $5$)
$\times$ ($10^7$, $10^8$, $10^9~\yr$) and are indicated by symbols.
The dashed gray line indicates the boundary of burst and non-burst
populations, i.e., e(b) and e(c), as defined in terms of \fOII\
equivalent width.  The dash-dotted grey line indicates the threshold
[$W_0(\Hd) > 4~\ang$] for e(a) population.  }
\label{fig.W0VsTimeExpoAndBrstMsun4}
\end{figure}

The \fOII\ emitters with e(a)-type spectrum make up a significant
fraction of the cluster \fOII\ emitters (\fig{fig.SCO2VsHd}), so we
place an emphasis on reproducing the e(a)-type spectrum for its
dominance in our sample.  We immediately excluded instantaneous burst
models, which were also modeled, as they generally fail to reproduce a
typical e(a) spectrum with strong \Hd\ absorption and \fOII\ emission
at the same time.  The values $\EBV = 0.5$ and $\EBVr = 0.44$ are
chosen as fiducial numbers representing our \fOII\ emitting sample.

In \fig{fig.W0VsTimeExpoAndBrstMsun4} we show the evolution of \fOII\
and \Hd\ equivalent widths of constant burst and
exponentially-declining SFH models.  The \Hd\ equivalent width cut for
e(a)-type galaxies is indicated by the horizontal dash-dotted lines.
Strong \Hd\ absorption is generally achieved by the models with a
shorter star-forming timescale, $\tau \la 100~\Myr$, and the phase can
continue up to a few \Gyr, roughly comparable to the lifetime of
intermediate age stellar population.  These short-$\tau$ galaxies are
generally classified into e(b)-type for up to $\sim 100~\Myr$ during
which they show strong \fOII\ emission ($<-40~\ang$), while their \Hd\
absorption has not had a time to grow as much.  After the vigorous
star formation, the short-$\tau$ galaxies go into a post-starburst
(k+a/a+k) phase rather quickly due to their fast fading \fOII\
emission.  Therefore short-$\tau$ models with the fiducial parameters
of \EBV\ and \EBVr\ are unlikely to explain the abundance of e(a)-type
galaxies in our sample (\fig{fig.SCO2VsHd}), since the short-$\tau$
models stay in the e(a) phase for no more than $\sim 100~\Myr$.

\begin{figure*}
\plotone{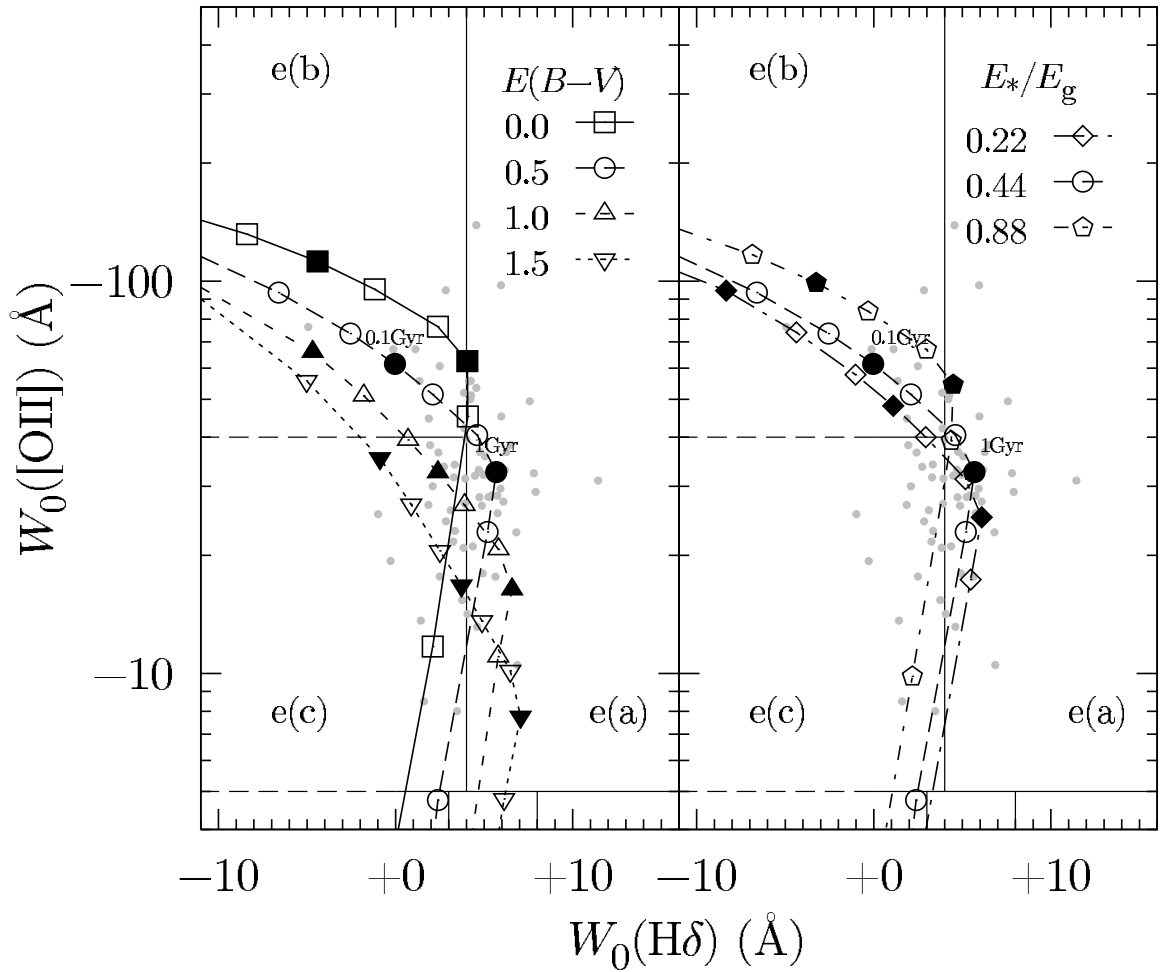}
\caption{ The effect of dust extinction on the observed spectral
indices of a fiducial model track (exponentially-declining SFH with
$\tau = 1~\Gyr$, $\EBV = 0.5$, and $\EBVr = 0.44$) indicated by
circles connected by dashed line.  (\emph{Left}) The output spectra
from the model were extincted by a different amount of \EBV\ at
constant $\EBVr = 0.44$.  (\emph{Right}) Stellar and nebular continua
were differentially extincted at a ratio given by \EBVr\ at constant
$\EBV = 0.5$.  Filled data points indicate models at $10~\Myr$,
$100~\Gyr$, and $1~\Gyr$.  Filled gray circles are observed cluster
\fOII\ emitters.  }
\label{fig.D99SCO2VsHdVaryEBV}
\end{figure*}

Strong \fOII\ emission \emph{and} \Hd\ absorption equivalent widths
are achieved for a more extended duration by the models with
star-forming timescales $\tau \sim 1~\Gyr$.  A major difference
between the constant burst and exponentially-declining SFHs with
similar $\tau$ is how star formation continues after $t\simeq\tau$
(which characterizes roughly the ``burst'' duration).  Since O- and
B-type stars can totally dominate the light from a galaxy when
present, the \Hd\ absorption line grows substantially only after a
significant amount of O- and B-type stars are gone.  Since star
formation totally ceases after a burst, the \fOII\ emission equivalent
width from constant burst models quickly goes below the boundary for
emission-line galaxies [$W_0(\fOII) >-5~\ang$] after a few \Myr.  On
the other hand, the $\tau \sim 1~\Gyr$ models with an
exponentially-declining SFH can spend up to several \Gyr\ within the
emission-line categories, which is desirable for reproducing a high
abundance of e(a) galaxies.

Now, \fig{fig.D99SCO2VsHdVaryEBV} illuminates the effect of dust
extinction in the \fOII-\Hd\ plane with a set of $\tau \sim 1~\Gyr$
exponentially-declining SFH models.  A range of reddening $\EBV\ =
0.0$--$1.5$ roughly spans the spread of \fOII\ emitters.  We point out
that the equivalent width is a measurement insensitive to the presence
of dust reddening when the continuum and line flux are both attenuated
by the same amount, in which case the model tracks do not deviate from
the dust-free, $\EBV=0$ model.  Therefore we introduced the effect of
differential extinction by varying $\EBVr$, the ratio of \EBV\ in
stellar to nebular component.  It is clear that differential
extinction is essential for pushing a model track into the e(a)
regime, as the \Hd\ equivalent width for the dust-free track turns
over at $\simeq 4~\ang$, barely crossing the boundary for the e(a)
phase; taking $\EBVr < 1$ extincts the nebular \Hd\ emission more than
the stellar continuum around it, enabling the equivalent width to grow
slightly larger in absorption (i.e., less emission filling).

However, the \fOII\ emission, being a bluer line, suffers even more
significantly from dust extinction as well.  This in turn leads to a
degeneracy between the age and the amount of reddening in the
\fOII-\Hd\ plane.  For example, a highly reddened $\EBV = 1.5$
starbursting galaxy at the age of $100~\Myr$ is hard to distinguish
from a $\EBV = 0.5$ galaxy at $\sim 3~\Gyr$, since they both would
have $W_0(\fOII) \simeq -11~\ang$ and $W_0(\Hd) \simeq 5~\ang$.  The
inferred age of star-forming activity can be quite different without
the knowledge of reddening for breaking such a degeneracy.

\subsection{Star Formation History by Composite Spectra}
\label{sec.SFHByCompositeSpectra}

We recall that it was largely the interpretation of low-\snr\ \Hd\
equivalent width measurements which led to the controversy regarding
the frequency of starbursts in cluster galaxies \citep{dre99,bal99}.
Although our analysis of e(a)-type galaxies in the \fOII-\Hd\ plane in
the previous section independently confirms the high abundance of
cluster galaxies with strong Balmer absorption lines, also reported
recently by \citet{dre04}, the relatively low spectral \snr\ of faint
\fOII\ emitters limits our ability to constrain the individual SFHs
with much confidence.  For our typical spectra, a $3\sigma$ detection
of \Hd\ absorption equivalent width at $W_0(\Hd) \sim 5~\ang$, for
example, needs per pixel $\snr \sim 10$ in the continuum, a
requirement not quite met by most of our spectra \citep{sat06}.  We
therefore supplement our analysis by composite spectra, which smear
out the SFHs of individual galaxies but would help us find the
characteristic SFHs of subsamples.

\begin{deluxetable}{cccc}
\tabletypesize{\footnotesize}
\tablecaption{$i$-$\log{\Sigma}$ Subsamples}
%\tablewidth{0pt}
\tablehead{
\colhead{Category} &
\colhead{$\log{\Sigma}$} &
\colhead{$i$} &
\colhead{$N$}
}
\startdata
HB  & $> 1.6       $ & $ < 21$ & 10 \\
TB  & $0.0$ -- $1.6$ & $ < 21$ &  3 \\
LB  & $       < 0.0$ & $ < 21$ &  6 \\
HM  & $> 1.6       $ & $21$ -- $23$ & 14 \\
TM  & $0.0$ -- $1.6$ & $21$ -- $23$ & 29 \\
LM  & $       < 0.0$ & $21$ -- $23$ &  6 \\
HF  & $> 1.6       $ & $ > 23  $ &  6 \\
TF  & $0.0$ -- $1.6$ & $ > 23  $ & 19 \\
LF  & $       < 0.0$ & $ > 23  $ & 11
\enddata
\label{tab.IMagSigmaSubsamples}
\end{deluxetable}

Our scheme for subsampling cluster \fOII\ emitters is motivated by the
observational evidence that the transformation of galaxies depends
both on their intrinsic property and on their local environment.  We
chose $i$ magnitudes and local galaxy densities $\Sigma$ to
characterize these attributes; \fig{fig.CoaddGridsIMagLogSigma} shows
the classification scheme, and \tab{tab.IMagSigmaSubsamples} shows the
number of objects within each subsample.  We effectively make
subsamples residing in the local environments characteristic of
low-density, field-like regions (L; $\log\Sigma < 0.0$), the
transitional zone (T; $0.0 < \log\Sigma < 1.6$), and the high-density
regions (H; $\log\Sigma > 1.6$).  The highest density cut was made
close to the \citetalias{kod01} threshold density.  Each density bin
is further divided into three $i$ magnitude bins: bright (B; $i <
21$), medium (M; $21 < i < 23$), and faint (F; $i > 23$).  The
subsampling yields composite spectra of quality ranging from per pixel
continuum $\snr \sim 3$ for faint subsamples to $\snr \sim 30$ for
bright subsamples.  For faint subsample this is still an improvement,
as most $i > 23$ \fOII\ emitters have $\snr < 2$.

The goal of this exercise is to study the SFHs inferred from
representative composite spectra of various cluster \fOII-emitting
subsamples by placing them in the \fOII-\Hd\ plane.  For this reason
the method for coadding spectra must not artificially change the
resulting \fOII\ and \Hd\ equivalent widths of composite spectrum,
i.e., continuum flux strength relative to the line flux must be scaled
properly at these lines.  Averaging spectra normalized to a common
mean flux density over a relatively featureless continuum region
\emph{for the entire spectrum} is not desirable, especially for
$\fOII\lam3727$, since the emission line appears within a sloping
continuum region of the Balmer break.  Instead we normalized spectra
to the continuum flux density at line center of each line of interest,
after the spectra are shifted to their rest frame wavelengths.  All
spectra are rebinned to a common dispersion at $1~\ang$ per pixel
before averaging to produce the final composite spectrum.  A potential
bias may be introduced to a composite spectrum depending on how each
spectrum is weighted or not weighted against others within the common
subsample.  We cannot weight by such quantities as inverse variance,
since it is strongly dependent on source luminosity, an intrinsic
property of subsample population which we wish to control.  In order
to treat each \fOII\ emitter within a subsample on equal footing, the
composite spectrum was computed without weighting.  The results are
also checked against the composite spectra made from medianing, which
is generally more robust against outliers.  We also excluded objects
identified to be AGN-like from our composite samples.

\begin{figure}
\plotone{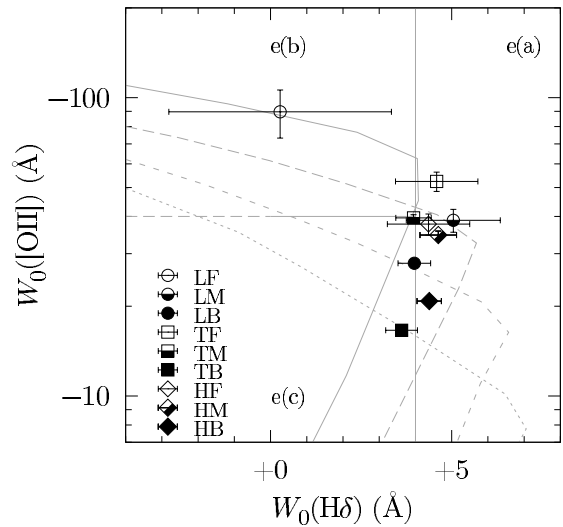}
\caption{ The \fOII\ and \Hd\ equivalent widths from the composite
spectra of $i$-$\Sigma$ subsamples.  The $1\sigma$ uncertainties were
computed formally from the sigma vectors of spectra to be coadded.
The gray lines indicate our fiducial emission galaxy models at $\EBV =
0.0$ (solid), $0.5$ (long-dashed), $1.0$ (short-dashed), and $1.5$
(dotted); see \fig{fig.D99SCO2VsHdVaryEBV}.  The uncertainties in
$W_0(\fOII)$ for some data points are smaller than their symbol size.
}
\label{fig.CoaddedOIIVsHd}
\end{figure}

In \fig{fig.CoaddedOIIVsHd} we see where the subsamples lie in the
\fOII-\Hd\ plane, overlayed on the evolutionary tracks of our fiducial
models for $\EBV = 0.0-1.5$.  Within similar local environments, the
\fOII\ equivalent width of \fOII\ emitters is strongly dependent on
their $i$ magnitude; luminous galaxies tend to have a weaker \fOII\
emission, in line with what we found in \citetalias{sat06}.
Remarkably, their \Hd\ equivalent widths are nearly constant at
$4-5~\ang$, and only the LF sample stands out with a small \Hd\
equivalent width in absorption ($\sim 0~\ang$).  The LF sample is
classified into e(b)-type, yet its \Hd\ measurement has a large
uncertainty; within $2\sigma$, the LF data point is consistent with
being e(a)-type.  While this is true, it is generally difficult for a
spectrum to have strong \fOII\ emission and \Hd\ absorption equivalent
widths at the same time because of emission filling at \Hd.  A very
strong \fOII\ emitter with $W_0(\fOII) \la -100~\ang$ would always
have \Hd\ equivalent width in emission, i.e., $W_0(\Hd) < 0~\ang$
(\fig{fig.D99SCO2VsHdVaryEBV}).  Therefore, with a reasonable
confidence in the very strong \fOII\ emission, it is likely that the
LF composite galaxy is a true e(b)-type galaxy.

It appears our model tracks with $\EBV = 0.0-1.5$ almost entirely
cover the range of measurements from the composite spectra.  While
this may indeed imply the importance of dust extinction, we expect the
$i$-$\Sigma$ sampling scheme to have degeneracy on the amount of dust
reddening in the \fOII-\Hd\ plane.  Composite sampling makes more
difficult to interpret the distribution seen in
\fig{fig.CoaddedOIIVsHd} in terms of the amount of reddening, since we
have seen there is a positive trend of \EBV\ on galaxy luminosity
\citepalias{sat06}.

\begin{figure}
\plotone{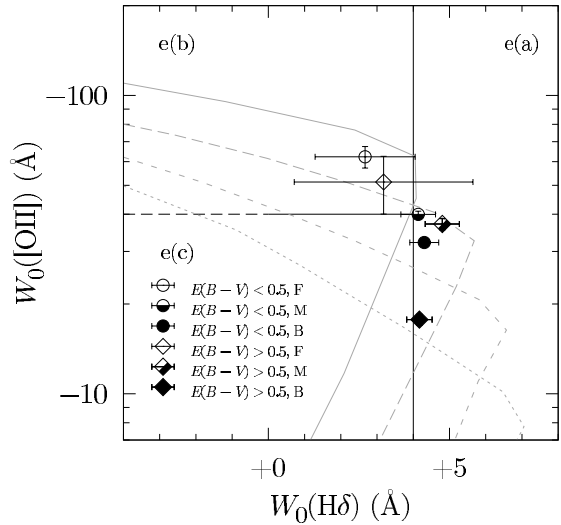}
\caption{ \fOII\ and \Hd\ equivalent widths of the composite spectra
from the $i$-\EBV\ subsamples (\sect{sec.SFHByCompositeSpectra}).  A
cut was made at $\EBV = 0.5$ to divide \fOII\ emitters into galaxies
with low and high reddening, and each subsample was further divided
into three luminosity bins with cuts at $i = 21$ and $23$.  The
uncertainties in $W_0(\fOII)$ for some data points are smaller than
their symbol size.  }
\label{fig.CoaddEBVIMagOnOIIVsHd}
\end{figure}

To better interpret the effect of reddening, we make another
subsampling based on \EBV\ in \fig{fig.CoaddEBVIMagOnOIIVsHd}.  On a
model evolutionary track, increasing \EBV\ has an effect of pushing
the entire track toward weaker \fOII-emitting e(a)-type galaxy
(\fig{fig.D99SCO2VsHdVaryEBV}).  If the dust reddening is the primary
parameter for the spread of \fOII\ emitters, resampling them based on
\EBV\ might form a one-parameter family of data points along the
evolutionary track on the \fOII-\Hd\ plane, but we see this is not the
case.  A cut was made at $\EBV = 0.5$ to divide \fOII\ emitters into
galaxies with low and high reddening, and each subsample was further
divided into three luminosity bins with cuts at $i = 21$ and $23$.
For the subsamples with similar luminosities, the effect of reddening
does not appear to be as strong as expected from the shift in the
model tracks arising from varying the amount of reddening alone,
except perhaps for the brightest subsamples.  The general trend of
decreasing strength of \fOII\ equivalent width with luminosity is
still observed.

The luminosity trend is particularly hard to interpret for moderate
\fOII\ emitters in the presence of dust, since evolutionary tracks
give rise to degeneracy in burst age and the amount of reddening.  For
example, in \fig{fig.CoaddEBVIMagOnOIIVsHd} the luminous (B), $\EBV >
0.5$ subsample lies close to the point at which the model tracks with
$\EBV = 0.5$ and $1.5$ intersect.  As far as the amount of reddening
is concerned, both the evolutionary tracks are very roughly consistent
with the measured \EBV, yet the inferred burst age is very different
depending on what the actual \EBV\ is---$\sim 100~\Myr$ for $\EBV =
1.5$ and $\sim 3~\Gyr$ for $\EBV = 0.5$.  Rather than constructing
composite samples, treating galaxies individually with distinct, more
precise reddening measurements would better constrain their SFHs in
this case, if the quality of spectra allows it.

\begin{figure}
\plotone{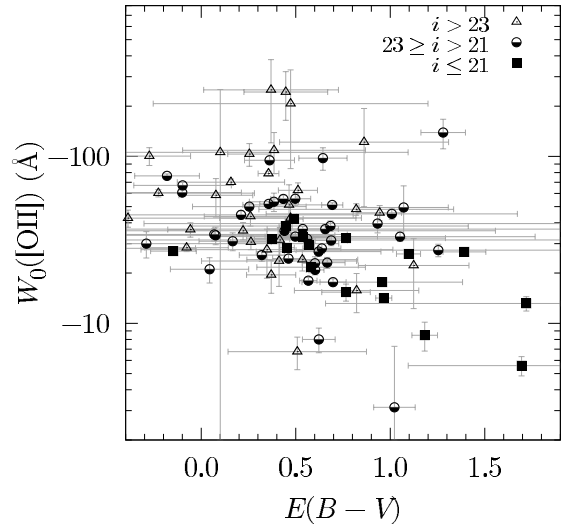}
\caption{ Rest-frame \fOII\ equivalent widths of cluster \fOII\
emitters as a function of their reddening \EBV.  Data points are
divided into low- ($i > 23$; open triangles), intermediate- ($23 \ge i
> 21$; half-filled circles), and high-luminosity ($i < 21$; filled
squares) sources.  Objects identified to be AGN-like are excluded from
the figure.  }
\label{fig.WO2VsEBV}
\end{figure}

To better elucidate the coupled effect of luminosity and reddening in
the \fOII-\Hd\ plane, in \fig{fig.WO2VsEBV} we relate the \fOII\
equivalent width, luminosity, and the amount of reddening.  We clearly
see that luminous ($i < 21$), weak-\fOII\ emitters [$W_0(\fOII) \sim
-10~\ang$] are highly reddened with $\EBV\sim 1$.  The bright
subsamples in \fig{fig.CoaddedOIIVsHd}, LB, TB, and HB, lie in the
region roughly bounded both by the tracks at the age of $\sim
1$--$2~\Gyr$ of low reddening $\EBV \simeq 0.0$--$0.5$ model galaxy
and the tracks at the age of $\sim 100~\Myr$ of high reddening $\EBV
\simeq 1.0$--$1.5$ model galaxy.  The observed high reddening suggests
the younger, highly reddened burst is a more natural interpretation of
luminous, weak-\fOII\ emitters.  In general, we see that large amount
of reddening can make young starburst galaxies appear in the same loci
as less reddened starburst at older burst age, and we tend to observe
them in the e(a)-type spectrum.  A high amount of reddening typically
observed in the cluster \fOII\ emitters [$\EBV \ga 0.5$] suggests that
younger burst ($\la 1~\Gyr$) is more likely to explain the dusty
\fOII-emitting population.

\section{Discussion}
\label{sec.Discussion}

Recent wide-field surveys of cluster galaxies have shown convincing
evidence that the transformation of galaxy properties starts well
beyond the virialized part of galaxy clusters, signaling the
importance of local environments for preprocessing of cluster galaxies
\citep[e.g.,][]{bal97, tre03, rin05, mor05, pim06}.  The star-forming
properties of \object{Abell 851} \fOII\ emitters have shown a strong
dependence on local environment as often reported in literature; the
\fOII\ emitter fraction depends strongly on local environments
\citep[e.g.,][]{bal04b, kod04, nak05}, while the strength of \fOII\
equivalent width is relatively independent of environment
\citep{bal04a, bal04b, kod04, rin05}.  These properties in \fOII\
emission are consistent with the interpretation that the mechanisms
that give rise to the suppression of star-forming activity either act
on very short timescales or are sufficiently rare, i.e., the abundance
of transient objects is relatively small.  Furthermore, there appears
a luminosity segregation of the \fOII-emitting populations in
dense environments---at the luminous end, there is an indication that
AGN activities could play a role in suppressing star-forming activity,
whereas cluster \fOII\ emitters are notably absent at the faint end.

\subsection{Timescales of Star-Forming Activity}
\label{sec.TimescalesOfStarFormingActivity}

We put some constraints on the timescales of star-forming activity in
the cluster \fOII\ emitters using a combination of stellar population
synthesis and nebular photoionization codes.
\figs{fig.D99SCO2VsHdVaryEBV}{fig.CoaddedOIIVsHd} show that the spread
of cluster \fOII-emitters in the \fOII-\Hd\ plane may be covered well
by exponentially-declining SFH models of $\tau \sim 1~\Gyr$, with
varying amount of reddening up to $\EBV \sim 1.5$.  A wide range of
reddening, however, gives rise to a degeneracy between the burst age
and reddening---a young, very dusty starburst and a less dusty
star-forming galaxy at an older stage of evolution can appear at a
similar locus in the \fOII-\Hd\ plane.

The composite spectra of the subsamples in the $i$-$\Sigma$ plane
indicate that the galaxy luminosity is a major factor contributing to
the spread of \fOII\ equivalent widths (\fig{fig.CoaddedOIIVsHd}).
This is expected, since there exists an empirical anti-correlation
between the \fOII\ equivalent width and galaxy luminosity
\citepalias{sat06}.  The interpretation of equivalent width is
somewhat complicated by the fact that it also is a measure of relative
fraction in young and old stellar populations; a galaxy with a small
vigorously star-forming regions spread within a much larger region of
old stellar population would have a weaker \fOII\ emission equivalent
width, for example.  Our spectral modeling assumed simplistic SFHs for
which $100\%$ of galaxy mass is produced in a single episode of star
formation.  A more reasonable scenario involves patches of young
star-forming regions embedded within older stellar populations.  In
such a case, both \fOII\ emission and \Hd\ absorption equivalent
widths from spatially integrated spectra would shift slightly toward
the origin ($W = 0~\ang$) in the \fOII-\Hd\ plane, in which the amount
of shift is sensitive to the relative abundances of young and old
stellar populations.  Albeit modeling extreme cases, it is still
encouraging that a simple treatment of differential extinction, as
suggested by \citet{pog01}, have reproduced the spread of observed
\fOII\ emitters in the \fOII-\Hd\ plane.

\fig{fig.SCO2VsHd} shows among the most abundant \fOII\ emitters in
\object{Abell 851} and in the field at high-$z$ are those with the
e(a)-type spectrum.  We re-emphasize that properly taking account for
the differential dust extinction in the stellar and nebular components
of a galaxy spectrum is essential for interpreting the e(a)-type
spectrum.  The dust-free model track barely reaches the \Hd\
absorption equivalent width strong enough to be considered an
e(a)-type in \fig{fig.D99SCO2VsHdVaryEBV}.  The identity of \Hd-strong
galaxies with moderate-to-strong \fOII\ emission has been somewhat
enigmatic, since emission-filling at \Hd\ absorption line can reduce
its apparent strength seen in absorption.  Although the nature of e(a)
galaxies and its role in cluster galaxy evolution remain debatable,
the original interpretation of e(a) galaxies as dusty starbursts by
\citet{pog99} has recently been given considerable boosts.  A study of
\SDSS\ galaxies at $z \sim 0.1$ by \citet{bal05} has concluded that
they are normal or dusty spiral galaxies with $\EBV \sim 0.7$,
assuming the Milky Way extinction of $R_V = 3.1$; a recent study of
mid-infrared sources also have supported the picture that e(a)
galaxies are dusty starbursting systems \citep{laf04}.

Although not a direct proxy for the presence of A-type stellar
population, a very large positive value of \Hd\ equivalent width must
indicate a dominant A-type star contribution over optical spectrum,
with the \Hd\ emission-filling potentially masking their presence by
reducing the equivalent width seen in absorption.  \Hd-strong galaxies
have been studied extensively in the context of
post-starburst/star-forming systems, since a substantial amount of
young O- and B-type stars have to be gone before the intermediate-age
stars can dominate the optical spectrum.  The simplest SFH that
produces a post-starburst/star-forming spectrum is a truncation SFH
after an episode of starburst or star formation.  For an e(a)-type
spectrum, however, some residual star-forming activity after a burst
is necessary to reproduce \fOII\ emission in dust-free models
\citep{pog99}.  Our modeling of e(a) spectra also does not favor
simple truncation of star forming activity
(\sect{sec.TimeEvolutionInTheOII-HdPlane}).

When differential dust reddening is accounted for, $\tau \sim 1~\Gyr$
models reproduce well the spread of cluster \fOII\ emitters in the
\fOII-\Hd\ plane.  That the timescale of $\tau \sim 1~\Gyr$ is of
order of A-type star lifetime is not a coincidence; large \Hd\
absorption equivalent widths simply require A- or early--F-type star
populations.  This brief window of order $1~\Gyr$ during which \Hd\
absorption is strong identifies either the recent cessation of
starburst or on-going starburst.  The former is the case for a
post-starburst galaxy since the nebular emission fades quickly due to
the absence of hot, young stars after a burst, and the latter is the
case for an e(a)-type galaxy since the nebular emission lines are
highly extinguished.  Hence strong \Hd\ absorption equivalent width
tends to signal the presence of starburst; high reddening further
implies the starburst could be younger than what dust-insensitive
indicators of star formation would suggest.

Since the inferred timescales, $\la 1~\Gyr$, of starburst are shorter
than cluster-crossing timescales (typically a few \Gyr), the higher
abundance of starbursting e(a) galaxies compared to the field fraction
strongly suggests the connection between enhanced star-forming
activity and infalling cluster substructures, as discussed in
\sect{sec.KinematicsOfTheClusterOIIEmitters}.

\subsection{The Cause of Enhanced Star-Forming Activity}
\label{sec.TheCauseOfEnhancedStarFormingActivity}

\fig{fig.FracSCVsZ} compares the cluster and field \fOII-emitting
samples, showing the e(a) abundance in \object{Abell 851} is roughly
at the same level as $z\sim1$ field galaxies among \fOII\ emitters.
Since the e(a) fraction significantly declines from $z\sim1$ to
$z\sim0.1$, it is likely that e(a) galaxies are fractionally more
abundant in \object{Abell 851} than in the field at a similar
redshift.  We argue that this enhancement of cluster e(a) population
is likely caused at the expense of continuously star-forming e(c)-type
galaxies.  First, our constraint on the burst age, $\la 1~\Gyr$, of
e(a) galaxies is consistent with the triggering of starburst upon
cluster infall, whose timescales are on the order of a few \Gyr.
Although the timescale argument does not necessarily prove the greater
frequency of starbursts upon infall [e.g., upon infall, e(c) galaxies
could drop out of the \fOII-emitting sample at a faster rate via
truncation of star formation than the e(a) production rate via
starburst], a higher burst fraction is consistent with post-starburst
(E+A) galaxies being more abundant \emph{within the total galaxy
population} in clusters than in the field at intermediate redshifts
\citep[][but see \citet{bal99}]{dre99, tra03, tra04}.  Second,
starburst galaxies almost certainly require gas-rich progenitors.  The
only class of galaxies with reasonably ample fuel supply is
star-forming e(c) galaxies.  Due to their short lifetime, e(b)
galaxies would not make a significant contribution.  Third, the
e(c)-to-e(a) transformation appears consistent with their morphology.
\citet{pog99} find e(a) and e(c) galaxies in $z \approx 0.4$ clusters
tend to show late-type disk morphology, but e(a) galaxies appear much
more likely to be associated with mergers or strong interactions.
\citet{bal05} also find field e(a) galaxies in $z \approx 0.1$ to show
disk-dominated morphology.  Furthermore, their distributions in the
\fOII-\Hd\ plane (\fig{fig.SCO2VsHd}) indicate e(a) galaxies are
neither rare or particularly special in their properties.  As clearly
seen in the \SDSS\ field sample, at least some galaxies in the e(a)
class simply constitute a high-\Hd\ extension of e(c) galaxies
following an evolutionary track for a small- to moderate-\EBV,
$\tau\sim$ several \Gyr\ exponentially-declining model, which is a
typical SFH for spiral galaxies.  Perhaps the e(a) galaxies of
\citet{bal05} constituted of a mixture of \Hd-strong, but otherwise
normal e(c)-type galaxies and real dusty disk galaxies observed during
an episode of starburst.  The set of evidence here naturally points to
the underlying connections between e(a) and e(c) galaxies---it is
reasonable that triggering of starbursts in gas-rich e(c) galaxies
gives rise to e(a) galaxies in these contexts.

The process of e(c) to e(a) transformation must induce increased
amount of dust in e(a) galaxies.  At high redshift beyond $z \ga 0.7$,
mergers are presumably more frequent, and infrared-luminous sources
appear to dominate star-forming activity \citep[e.g.,][]{lef05},
implying the importance of star-forming activity obscured by dust.
These observations indicate the apparent connection between
dynamically active environment and its effect on inducing dusty
starburst.  Although major mergers are not likely to result in disk
systems, merger-induced dusty starburst has been proposed as a
mechanism for reproducing e(a) spectrum \citep{shi00}; dust plays an
essential role in prolonging the e(a) spectrum phase by extinguishing
emission lines from star-forming regions.  It is known that most
vigorously star-forming galaxies emit strongly in infrared through
dust reprocessing of UV light from young stars.  Thus it is not
surprising that e(a) galaxies are dusty, if they are in fact starburst
galaxies.  Furthermore, the selectively higher extinction of light
from very young stellar population \citep{pog01} would tend to yield
high reddening values, if the nebular lines originating from vigorous
star-forming activity can be measured at all.  Our spectral
evolutionary tracks assumed constant reddening \EBV\ throughout the
lifetime of a starburst galaxy.  If, instead, the amount of extinction
has a decreasing trend with the age of the stellar population sampled
for observation, the fact that e(a) galaxies are dustier just reflects
the significant presence of very young stellar populations on top of
underlying older stellar populations.  Therefore the high reddening
values favor a picture of e(a) galaxies as starbursts by inferring the
presence of very young stellar populations but may not help us
constrain one mechanism over others as to what actually causes
starbursts, i.e., dusty starbursts do not \emph{have} to be caused by
mergers or interactions.

\begin{figure}
\plotone{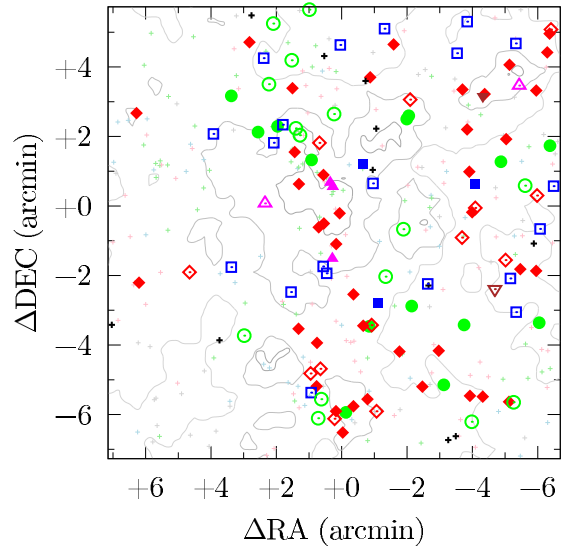}
\caption{ Spatial distribution of cluster \fOII\ emitters in terms of
their \citetalias{dre99} spectral class: e(b) (blue squares), e(c)
(green circles), e(a) (red diamonds), k (brown inverted triangles),
and k+a/a+k (magenta triangles).  Black crosses denote \fOII\ emitters
for which a spectral class cannot be determined.  Filled symbols
denote the spectral class derived from better than $2\sigma$
measurements of line indices, whereas open symbols denote that from
$<2\sigma$ measurements.  Smaller, slightly faded crosses are
\citetalias{mlf00} \fOII\ emission-line candidates not observed
spectroscopically; these are \gi\ color-coded by blue ($1<\gi$), green
($1<\gi<2$), red ($\gi>2$), and gray (the uncertainty in \gi\ is
$>0.5$).  The contours are the same as in
\fig{fig.VelocitySpatialDist}.  [\emph{See the electronic edition of
the paper for a color version of this figure.}]  }
\label{fig.D99SCSpatDist}
\end{figure}

The question yet to be answered clearly is what physical mechanisms
are causing the cluster \fOII\ emitters to go through starbursts.  To
address this, we see in \fig{fig.D99SCSpatDist} the spatial
distribution of cluster \fOII\ emitters with their \citetalias{dre99}
spectral class.  A remarkable observation is the high degree of
clustering of e(a) galaxies near or inside the infalling groups, i.e.,
the southern and north-western clumps of galaxy overdensities, and the
periphery of cluster central region.  The apparent increase of
galaxies showing starburst signatures in their spectra near the
\citetalias{kod01} threshold density (\fig{fig.SCVsLogSigma}) offers
another evidence for such a trend.  In galaxy groups beyond the
cluster virial region, both major mergers \citep{ick85, mih95, bek98}
and galaxy harassment \citep{moo99} are probable, but the effect of
intracluster medium is expected to be less effective.  Therefore
galaxy-galaxy interactions offer a natural explanation for the
starburst triggering mechanisms in those environments.  The best way
to confirm this hypothesis is to look at their morphology for signs of
interactions.  Some high-resolution Wide Field Planetary Camera 2
(WFPC2) images from \emph{Hubble Space Telescope} (\HST) exist for
\object{Abell 851}, but unfortunately the infalling substructures seen
in \fig{fig.D99SCSpatDist} are outside of their coverage.

\begin{deluxetable*}{ccccccccl}
\tabletypesize{\footnotesize}
\tablecaption{Cluster \fOII\ Emitters with \HST\ WFPC2 Morphology}
\tablewidth{0pt}
\tablehead{
\colhead{\citetalias{mlf00}} &
\colhead{$\Delta$RA\tablenotemark{b}} &
\colhead{$\Delta$DEC\tablenotemark{b}} &
\colhead{\citetalias{dre99}} &
\colhead{S97} &
\colhead{Hubble} &
\colhead{D\tablenotemark{f}} &
\colhead{Interpretation\tablenotemark{g}} &
\colhead{Comment} \\
\colhead{ID\tablenotemark{a}} &
\colhead{($'$)} &
\colhead{($'$)} &
\colhead{Class\tablenotemark{c}} &
\colhead{ID\tablenotemark{d}} &
\colhead{Type\tablenotemark{e}} &
\colhead{} &
\colhead{} &
\colhead{}
}
\startdata
1448 & $+0.76$ & $-3.94$ & e(a) & \emph{314} & Sbc & 2 & Tidal feature? &
Face-on disk or disturbed spheroid?  Shells or arms? \\
1640 & $+1.32$ & $-3.53$ & e(a) & \emph{2022} & Sb & 0 & &
Multi-armed disk galaxy; bright central disk, small bulge \\
2095 & $+1.55$ & $-2.48$ & \emph{e(b)} & \emph{499} & Sc & 1 & &
Small, highly inclined disk; slight structure visible \\
2303 & $+0.46$ & $-1.94$ & \emph{e(b)} & \emph{N/A}\tablenotemark{h}  &  &  & &
\\
2978 & $+0.70$ & $-0.61$ & e(a) & 2016 & E & 0 & &
\\
3005 & $+0.55$ & $-0.50$ & e(a) & 620 & Sb/S0? & 1 & Tidal feature? &
Small, high SB spheroid; featuring disk w/arms, or tidal \\
3215 & $+0.06$ & $-0.21$ & e(a) & 497 & Sbc & 2 & Merger &
Visually disturbed, high SB spiral; probably merger with tail at 11 \\
3568 & $+1.31$ & $+0.63$ & e(a) & 34 & SBab & 0 & &
High SB bulge or nucleus; weak arms from bar \\
3600 & $+0.27$ & $+0.55$ & k+a & 122 & Sd & 3 & Merger &
Disk system with tails; bright blobs---likely merger \\
3604 & $+0.35$ & $+0.67$ & a+k & 36 & Sd & 2 & Merger &
Very high SB, disky object w/ 2 blobs and tail
\enddata

\tablenotetext{a}{ \citetalias{mlf00} object ID from their \fOII\
emission-line search. }

\tablenotetext{b}{ Relative to \clustercenter. }

\tablenotetext{c}{ \citetalias{dre99} spectral class derived from our
measurements.  The classification derived from $<2\sigma$ measurements
are italicized. }

\tablenotetext{d}{ Cross-identification with \citet{sma97}.  The IDs
from the pointing outside the cluster center, i.e., Cl0939+47 Field 2
in their catalog, are italicized. }

\tablenotetext{e}{ The standard Hubble classification scheme (E, S0,
Sa, Sb, etc.). }

\tablenotetext{f}{ Disturbance class: 0, normal; 1, moderate
asymmetry; 2, strong asymmetry; 3, moderate distortion; 4, strong
distortion. }

\tablenotetext{g}{ Interpretation of disturbance classes, roughly
divided into (M) merger, (I) tidal interaction with neighbor, (T)
tidal feature, (C) chaotic, and (!) remarkable. }

\tablenotetext{h}{ No matching object found in the morphology
catalog. }

\tablecomments{ First four galaxies are in Field~2 of the
\citet{sma97} survey which sampled galaxies in the ``outskirts'' of
Abell~851, and the latter six are from the cluster central region from
the same survey.  }

\label{tab.Morph}
\end{deluxetable*}

Turning our attention to the cluster central region, we see e(a)
galaxies residing mostly in the \emph{periphery} of the highest
galaxy-density contour (\fig{fig.D99SCSpatDist}).  Although some of
these \fOII\ emitters could be 2D-projected foreground/background
galaxies in moderately dense infall group environments along the
filament outside the core, the \fOII\ emitters are mostly seen in the
lowest velocity bin in \fig{fig.VelocitySpatialDist}, indicating their
radial velocities are relatively close to the cluster systemic.  In
the core environment, both the intracluster gas and gravitational
potential of the cluster can interact with cluster galaxies.  Major
galaxy-galaxy mergers are expected to be very rare, since the deep
gravitational potential well of cluster means that the galaxies close
to the cluster dynamical center are moving very fast, reducing the
probability for a low-speed encounter with another galaxy.  Yet galaxy
harassment, in the sense of high-speed encounters between galaxies,
would still be effective in the central region.  The morphological
analysis using \HST\ has been done for many galaxies in the central
region of \object{Abell 851} by well-trained morphologists.  In
\tab{tab.Morph} we list a common subset from the catalog by
\citet{sma97} and our \fOII\ emitters.  Of particular interest here
are those objects found within the cluster central region (the latter
six in the list).  Four out of six galaxies have their disturbance
class greater than 0, meaning there is some morphological sign of
interaction in those galaxies.  Of these, three are presumably
mergers, although we are not sure whether the \citeauthor{sma97}
interpretations distinguish between high-speed (i.e., harassment) and
low-speed galaxy-galaxy encounters (i.e., major/minor mergers).  As
the comment indicates the remaining one is not a clear-cut case, and
the cause of its disturbance is not clear.

In any case, the existing morphological data in \object{Abell 851},
though limited in quantity, indicate a high fraction of the cluster
e(a) galaxies are involved in interaction.  As discussed earlier,
\citet{pog99} already noted a higher incidence of mergers or
interactions in e(a) galaxies compared to e(c) galaxies from the full
\citeauthor{sma97} morphology catalog of intermediate redshift
clusters.  We further note that most of the \fOII\ emitters in the
central region are morphologically classified into disk galaxies, yet
their spectral signature implies a current or recent starburst.  In
the presence of high reddening, their optical colors and the strength
of \fOII\ emission may not give an obvious indication of starburst, so
the apparently ``normal'' spiral galaxies in the cluster center, which
are photometrically more akin to field spirals than cluster spirals in
nearby clusters \citep{and97}, could be going through bursts.

Of course it is of no surprise that starbursts are induced by some
form of interaction; galaxy groups provide ideal environments for
merger-induced star-forming activity.  We now speculate if the
occurrence of starbursts could be enhanced specifically in cluster
environments.  X-ray observations show that \object{Abell 851} is
going through an on-going cluster-scale merger process, in which two
X-ray emitting cores merging in about a few hundred \Myr\
\citep{schi96, schi98, def03}.  Recent simulations have shown that
galaxy clusters in \LCDM\ cosmology have two distinct phases in mass
accretion history, merger-dominated and accretion-dominated phases
\citep{zha03}.  Major cluster mergers, involving almost doubling their
total mass, is not rare at the redshift of \object{Abell 851}
\citep{tas04}.  Inferred star-forming property of cluster galaxies
would show differences in each phase, adding possible complications to
a pure infall scenario.  The apparently complex spatial as well as
velocity distributions of \fOII\ emitters in \object{Abell 851}
perhaps reflects such cluster-scale mixing of galaxy populations.
Applying dynamical methods to such a dynamically active cluster poses
difficulties as well; velocity dispersions are not a good indicator of
cluster properties for dynamically active clusters going through
mergers \citep{cro95}.\footnote{As a result, the estimates for the
virial radius of \object{Abell 851} vary widely, and its unique
dynamical center can not be rigorously determined.  This is a major
reason why we have stayed clear of analysis using cluster-centric
radii.}

On the flip side of the complexity, there are some evidence that
dynamically active clusters host a larger number of star-forming
galaxies and AGNs \citep{lot03, mil03, rak05}.  As mentioned
previously the existence of old galaxy population tracing clumpy
distribution of substructures along filaments is generally confirmed
in wide-field survey of \object{Abell 851} using both photometric
redshift method \citepalias{kod01} and cluster red-sequence technique
\citep{kod05}.  The radial velocity distributions show weak, yet
positive evidence for kinematic segregation between the cluster \fOII\
emitters and \citetalias{dre99} k-type galaxies, when the samples are
restricted within the core region
(\fig{sec.KinematicsOfTheClusterOIIEmitters}).  Their spatial
distributions, again restricted within the central region, clearly
show an abundance of old galaxies where the cluster \fOII\ emitters
only sparsely exist.  Hence it is natural to interpret many \fOII\
emitters to be relatively new, recently accreted galaxies onto the
cluster potential which already hosts many evolved galaxies.  The
non-Gaussian velocity distribution of cluster \fOII\ emitters also
provides evidence that they represent a population that has not been
virialized in the cluster potential.  In general, it is unlikely that
red, gas-poor galaxies such as cluster ellipticals get rekindled in
starbursts, but a dominant contribution to cluster star-forming
activity can easily come from gas-rich spiral galaxies infalling onto
a cluster.  Hence, the enhanced star-forming activity in the infall
region, like that seen in the southern substructures in
\fig{fig.D99SCSpatDist}, probably is due to an increased frequency of
galaxy-galaxy interactions between quiescent yet gas-rich galaxies, as
the infalling groups themselves interact with nearby groups on the way
to the cluster center.  This is consistent with the high abundance of
\fOII\ emitters \emph{between} the apparent infalling groups in the
southern filament.

Yet clusters offer means of interactions other than the ones between
galaxies---the dense cluster gas in the core could ram-strip galaxies
of their fuel supply, and the steep cluster gravitational potential
could tidally interact with infalling galaxies.  In
\object{CL0024+1652} at $z \approx 0.4$, \citet{mor05} found that
morphologically early-type galaxies with large mass showed enhanced
Balmer absorption lines within the cluster virial radius, while some
fainter counterparts close to the virial radius actually showed \fOII\
emission.  These authors rule out merger-induced star formation after
inspection of their morphology (though they mention harassment is a
potential triggering mechanism).  In local Coma cluster, \citet{pog04}
recently found an abundance of young post-starburst galaxies near the
edges of X-ray substructures, indicating the interaction between
cluster galaxies and intracluster medium could trigger star-forming
activity.  These observations imply the importance of interactions
between infalling galaxies and environments specific to clusters.  The
X-ray observation by \citet{def03} was probably not deep enough to
trace the cluster gas much beyond the very center of \object{Abell
851} (\fig{fig.Core}).

To summarize, in view of highly dynamically active state of
\object{Abell 851}, the enhancement of star-forming activity seen in
this cluster may be the result of accelerated gas consumption within
infalling gas-rich galaxies driven by the galaxy-galaxy interactions
in infalling group environments.  As a favored triggering mechanism of
starbursts, galaxy-galaxy harassment, as opposed to a major merger,
appears consistent with preserving the disk morphology of cluster
\fOII\ emitters, although the direct confirmation with our
\fOII-emitting sample is limited only to the cluster central region
due to the limited coverage of \HST\ observations.  For the rest with
subtle or no morphological indication of interaction in the cluster
central region, we can only speculate if an increased inhomogeneity in
the distribution of intracluster gas/potential makes triggering
star-forming activity more effective, such as in shocking the
interstellar medium of galaxies.

\subsection{Possible Role of AGNs as Firemen}

\begin{figure}
\plotone{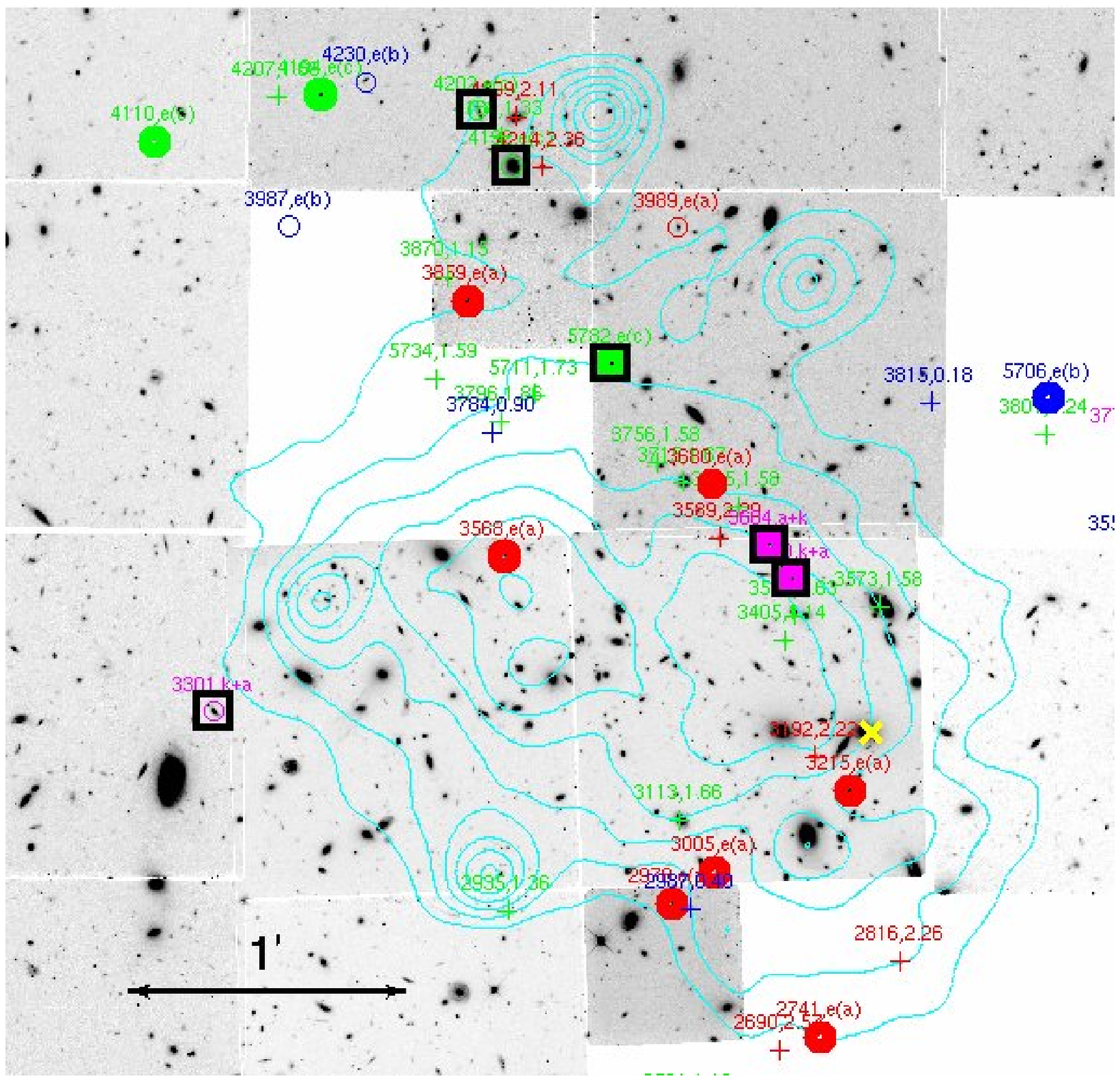}
\caption{ The distribution of cluster \fOII\ emitters in the central
region of Abell 851 with the \HST\ WFPC2 images.  The north is up, and
the east is in the left.  The yellow cross denotes \clustercenter,
from which $\Delta{\rm RA}$ and $\Delta{\rm DEC}$ are measured in
\fig{fig.VelocitySpatialDist}.  The overlayed contours in cyan are
X-ray emission map from \citet{def03}.  Those galaxies identified to
be AGN-like are marked by black squares; otherwise the colors of the
symbols denote \citetalias{dre99} spectral classes as in
\fig{fig.D99SCSpatDist}.  [\emph{See the electronic edition of the
paper for a color version of this figure.}]  }
\label{fig.Core}
\end{figure}

Some optically-luminous \fOII\ emitters in the cluster central region
shows a glimpse of the fate of cluster star-forming galaxies.  To see
how the \fOII\ emitters are distributed relative to the intracluster
gas, in \fig{fig.Core} we overlay X-ray emission contours, kindly
provided by B. De~Filippis.  We see very few \fOII\ emitters in the
region where the X-ray emission is very strong, although galaxies of
the types associated with old stellar population are abundant (see
\fig{fig.VelocitySpatialDist}).  More remarkably, the \fOII\ emitters
classified as AGN-like (in black squares) appear to align themselves
in a preferred direction from the south-west to the north-east.
Whether this preferred axis has any physical significance is not
obvious both from the figure and from the X-ray temperature map
presented by \citet{def03}.  Nonetheless we note that the cluster
central field presented in \fig{fig.Core} is a small fraction of the
total survey area for our \fOII\ emission search but hosts six of the
seven AGN-like galaxies in our sample; in comparison, only $20$ out of
$111$ confirmed cluster \fOII\ emitters and $186$ out of $808$
\citetalias{kod01} cluster members are in the region shown in
\fig{fig.Core}.

We reiterate that our AGN identification is not very complete, i.e.,
the line-ratio diagnostics could be done only for a subset of all the
cluster \fOII\ emitters.  Furthermore, none of the AGN-like galaxies
are spatially coincident with X-ray point sources identified by
\citet{def03}; only one $\gi > 2$ \fOII\ emission candidate (which was
not followed up spectroscopically) from the \citetalias{mlf00}
narrowband survey coincided with the point source \#8 in
\citeauthor{def03}\ The optical diagnostics are often inadequate for
secure AGN identification; \citet{mar02}, for example, have reported
in \object{Abell 2104} the optically-selected AGNs fraction is by a
factor of $5$ lower than that of the X-ray--selected---some
X-ray--selected AGNs do not even show strong emission lines.  On the
other hand, a high X-ray column can give rise to X-ray--obscured,
optically-identified AGNs \citep{bra04}.  Based on the fluxes of the
X-ray point sources detected in their imaging, the \emph{XMM} exposure
time of $51~{\rm ks}$ should have easily detected AGNs at the flux
level of $f_X(0.3-10.0~{\rm keV}) \ga 10^{-14}~\erg~\s^{-1}~\cm^{-2}$.
To our knowledge it is not clear whether a reasonable level of high
X-ray column can obscure AGNs to put our galaxies below this level of
detection.  Nonetheless the optical line ratio diagnostics does
suggest a possibility that these AGN-like galaxies could harbor
low-luminosity AGNs.

If these are in fact AGNs, a possible clue may be provided by a recent
study by \citet{yan05}, reporting that the \fOII\ emitters in red
sequence and post-starburst galaxies in their low-$z$ \SDSS\ sample
seem to host AGNs/LINERs.  In an independent study, Tremonti et
al. (2006, private communication) similarly finds signs of AGN
activity in their \SDSS\ post-starburst galaxy sample.  Spectral
decomposition of an \SDSS\ AGN sample by \citet{van06} showed
post-starburst activities to be much more common in broad-line AGNs.
Interestingly, three of the AGN-like galaxies in our sample are
classified into k+a/a+k, a post-starburst type, in the
\citetalias{dre99} spectral classification scheme, although the fact
that they made into our \fOII-selected sample means they have residual
\fOII\ emission.  We also find most of the AGN-like galaxies in our
sample reside in local environments whose galaxy densities are higher
than the \citetalias{kod01} threshold above which cluster galaxy
colors are typical of red, non--star-forming population (although we
must point out the \citetalias{kod01} threshold characterizes the
dense cluster central region, as well as infalling substructures where
we find little evidence for AGN-like galaxies).  Furthermore, their
location on the color-magnitude diagram indicates they are among the
most luminous \fOII\ emitters, but their \gi\ color is either
indistinguishable from the rest of the \fOII-emitting population or
consistent with being on the red sequence (\fig{fig.GIClrMag}).  With
the indications that the cluster red sequence building up from the
luminous end in a downsizing fashion \citep{del04, tan05}, there is a
curious connection between AGNs as a mechanism for shutting off star
formation and the property of host galaxies.  Hence the distribution
of AGN-like \fOII\ emitters in the color-magnitude diagram, along with
their relatively smaller \fOII\ emission equivalent widths
\citepalias{sat06}, is consistent with the scenario in which the
star-forming activities within luminous galaxies are shut off by AGNs.
Similar observations of cluster star-forming galaxies over a wide
range of redshifts could provide evidence for or against the role of
AGNs in causing the downsizing trend, where quenching of star
formation progressively move toward lower luminosity systems at lower
redshifts.

Yet, it is a mystery as to why AGN-like galaxies are not seen in the
\fOII\ emitters outside the cluster central region.  The
\citetalias{kod01} threshold effect, if real, implies that galaxies
can transform from star-forming to non--star-forming in group
environments in infall regions.  Six out of seven AGN-like galaxies in
our sample are in high-density environments above the
\citetalias{kod01} threshold but are preferentially located in the
cluster central region.  \citet{rud05} have generally found an excess
of X-ray--selected AGNs in the core region in both relaxed and
morphologically-disturbed clusters (plus another milder peak near
virial radius in relaxed clusters).  These observations might suggest
the significance of the locations of galaxies within clusters, as well
as the effect of local environments, in transforming the star-forming
properties of cluster galaxies.

There also is an ambiguity as to whether their broadband $i$
magnitudes are truly representative of the host galaxies.  If AGNs are
active, it is likely that they make a significant contribution to the
continuum light in the optical, making host galaxies brighter than
they really are.  That the most AGN-like galaxies in our sample are
among the brightest \fOII\ emitters might simply be a result of
additional contributions to the continuum from AGN activities.

In view of the general difficulty in systematic AGN selection, the
AGN-like galaxies in our \fOII-emitting sample do not offer direct
evidence for the connection between quenching of star formation and
AGN activity.  Yet, the observed trends discussed here are all
consistent with AGNs playing some role in the production of bright
cluster ellipticals with little star-forming activity.  In any case, a
more robust AGN identification and follow-up studies of these objects
would be necessary to confirm this hypothesis.

\subsection{Interpretation of Faint \fOII\ Emitters}

The analysis of composite spectra based upon sampling on the
$i$-$\Sigma$ plane has shown relatively little variation with local
galaxy density, except for the faintest \fOII\ emitters residing in
the low-density environment, i.e., the LF subsample.  The \fOII\
emitters in the other subsamples are mostly consistent with the
e(a)-type and typically show higher \Hd\ absorption equivalent widths
at $\sim 4$--$5~\ang$, with relatively little scatter.  The faint
\fOII\ emitters in low-density environments show the \fOII\ and \Hd\
equivalent widths typical of e(b) spectrum indicating a very young,
pure starburst.

\nbody\ simulations by \citet{moo99} demonstrated that the response of
a spiral galaxy to the tidal disruption from the cluster potential and
galaxy-galaxy interactions---galaxy harassment---varies depending on
the concentration of the mass distribution and the disk scale length.
Low surface brightness galaxies respond dramatically to the influence
of both the local and global variations in the gravitational
potential, while high surface brightness galaxies remain stable.  We
lack morphological information to see how $i$ magnitude and mass
concentration are related systematically in our sample, though faint
$\gi < 1$ \fOII\ emitters generally have small isophotal area,
implying their dwarf identity \citepalias{mlf00}.  The relative
depletion of HF \fOII\ emitters in \fig{fig.CoaddGridsIMagLogSigma}
could be evidence for destruction of star-forming dwarf galaxies in
dense regions due to galaxy harassment.  On the other hand, such an
absence of faint star-forming population does not have to indicate the
total absence of galaxies in the particular subset.  The depletion of
faint \fOII\ emitters in dense environments may just be a reflection
of highly suppressed star-forming activity in dwarf galaxies.  The
dense intracluster medium can provide effective gas-removal
mechanisms, such as ram-pressure stripping, which is also a very
likely scenario for dwarfs with a more diffuse matter distribution.
In this case galaxies are not destroyed but could simply fade out of
$i$ band flux limit as star formation ceases.

The contribution of dwarf galaxies to the total galaxy population has
been studied in terms of galaxy luminosity functions
\citep[e.g.,][]{dep98a, dep98b, tre02, val04, pop05, pop06}.  Up to
and below the magnitude limit relevant for the \fOII\ emitters studied
in this paper, there are evidence for a steeper faint-end slope in
nearby clusters than that of field luminosity functions, and a higher
abundance of dwarf galaxies in clusters has been proposed as the cause
of such steep cluster luminosity functions \citep[e.g.,][]{dep98a,
dep98b}.  The detection of such a trend seems not universal, however,
and there is even a suggestion that sample selection biases could
cause a steeper faint-end slope \citep{val04}.  Although it is still
not clear how the general form of faint-end luminosity function
differs systematically between clusters and field regions,
\emph{type-specific} luminosity functions appear to depend strongly on
environment regardless of the precise form of luminosity function for
the entire population: The brighter end is dominated by
non--star-forming population, while the faint-end is very sensitive to
the relative abundance of both star-forming and non--star-forming
populations \citep{got02, dep03}.

Two major surveys, \SDSS\ \citep{got02} and \TwodFGRS\ \citep{dep03},
disagree in their measurement of the faint-end slope of the cluster
luminosity function---flatter in \SDSS\ and only slightly steeper in
\TwodFGRS.  As far as the interpretation of where the star-forming
dwarfs in high-density environments go, the behavior of the faint-end
slope makes a profound difference.  If the disappearance of
star-forming faint dwarfs is simply due to their fading as in
stripping of gas supply, an increase in the number of
non--star-forming dwarfs would match the decrease in the star-forming
population, and the shape of overall luminosity function would not
change as much, consistent with \TwodFGRS, finding non--star-forming
population to increase and make the faint-end slope steeper (less
complications arising from flux limits being sensitive to galaxies
with different stellar populations).  On the other hand, galaxy
harassment can either destroy galaxies as in tidal debris
\citep{moo99} or \emph{create} even smaller galaxies as in tidal dwarf
galaxies \citep{pop06}.  Hence the behavior of the faint-end slope of
luminosity function alone would still not distinguish these dwarf
galaxy production models.  Our sample is also insensitive to the
surviving dwarf galaxies in the cluster central region, since they are
likely to be stripped of their gas supply \citep{mor00} and show no
star formation.

Yet our data could offer insight as to at which point of cluster
formation history the abundant cluster dwarf population is really
produced.  Since the metal enrichment ceases in such gas-poor dwarfs,
comparing the gas-phase metallicities of dwarf galaxies in
\object{Abell 851} and local clusters may tell us if gas-rich dwarfs
in $z = 0.4$ clusters could be the progenitors of the abundant
gas-poor dwarf galaxies in local clusters.  We wish to present the
metallicity analysis in our forth-coming paper of the series.

In short, our finding in \object{Abell 851} merely shows high-density
environments to be hostile to dwarf starburst galaxies.  The reported
increase in non--star-forming dwarfs in cluster central region
\citep{tho93, dep03, pop06} does not necessarily prefer one
interpretation over another as to the fate of infalling star-forming
dwarfs.  The forth-coming metallicity analysis may provide some clues
for addressing the progenitor-offspring issues surrounding cluster
dwarf galaxies.

\section{Summary}
\label{sec.Summary}

We have reported on a spectroscopic follow-up study of
narrowband-selected \fOII\ emitters in \object{Abell 851} from the
\citetalias{mlf00} catalog, focusing on their local environments,
kinematics, and SFHs.  Along with similar studies
\citep[e.g.,][]{bal04a, bal04b, kod04, rin05}, the cluster \fOII\
emitters reaffirm the fundamental importance of local environments:
The fraction of galaxies with \fOII\ emission is a strong function of
local galaxy density where few \fOII\ emitters are found in dense
environments, yet the strength of \fOII\ equivalent width does not
vary strongly with local galaxy density.  The trends are consistent
with the mechanisms that halt star-forming activity on short
timescales or are sufficiently rare.  In high-density environments,
the cluster \fOII\ emitters appear to show distinct behaviors
depending on their luminosity: For optically-luminous \fOII\ emitters
in the cluster central region, there is some weak evidence for AGNs
playing a role in suppressing star formation in the host galaxies.  On
the fainter end, we note the lack of faint \fOII\ emitters in dense
environments; the destruction of dwarf galaxies in galaxy harassment
\citep{moo99} and/or simple fading of emission lines due to quenching
of star formation in ram-stripped dwarfs \citep{mor00} can cause such
a deficit.  The luminosity-dependent behaviors of star-forming
galaxies strongly suggest normal disk galaxies, which are
predominantly e(a) galaxies, and dwarf galaxies, which are mostly pure
bursting e(b) galaxies, respond quite differently within dense cluster
environments.

The cluster \fOII\ emitters have shown the kinematic properties
consistent with a non-virialized population; their velocity
distribution deviates significantly from a Gaussian.  The skewed
distribution toward higher velocities is also consistent with a high
fraction of the \fOII\ emitters being at first infall, as opposed to a
backsplash population which may have gone through the cluster central
region in the past \citep[e.g.,][]{gil05}.  We observe a greater
fraction of cluster \fOII\ emitters with high radial velocities,
compared to the emission-line galaxies in local clusters
\citep{rin05,pim06}.  In addition, their spatial distribution is
highly asymmetrical with respect to the cluster center.  These provide
evidence for a signification fraction of the cluster \fOII\ emitters
constituting an infalling population.

The spectral evolution modeling and the distribution of the cluster
\fOII\ emitters in the \fOII-\Hd\ plane highlight a dominant presence
of dusty starburst galaxies [i.e., e(a)-type in the \citetalias{dre99}
scheme].  A high abundance of e(a) galaxies in clusters was already
remarked by \citet{pog99}, whose subsequent studies identified them as
dusty starbursts, primarily caused by selective extinction of the
light from very young stellar population \citep[e.g.,][]{pog00,
pog01}.  A combination of stellar population synthesis and nebular
photoionization codes favors models with star-forming timescales of
order $\la 1~\Gyr$ for dusty galaxies, $\EBV \ga 0.5$, for covering
the distribution of cluster \fOII\ emitters in the \fOII-\Hd\
equivalent width plane.  The range of timescales, along with a higher
occurrence of e(a) galaxies in \object{Abell 851} compared to the
field, is consistent with scenarios in which triggering of the
starbursts is cluster-related.  Existing morphological studies
\citep{sma97, pog99} further provide evidence for e(a) galaxies
arising at the expense of regular star-forming galaxies [i.e.,
e(c)-type] by preserving their disk-like morphology.  The redshift
trend of the relative fractions of \fOII\ emitters in various
\citetalias{dre99} spectral classes also suggest a higher incidence of
e(a) galaxies in \object{Abell 851} compared to the field at a similar
redshift.

The enhanced star-forming activities observed in \object{Abell 851}
may have some connections to the highly dynamically active state of
the cluster.  We specifically note a high abundance of \fOII\ emitters
\emph{between} two infalling groups located in the southern filament
of \object{Abell 851}.  The proximity of two groups during the process
of an accretion onto the cluster may have caused the galaxies to
interact at a higher frequency, although we lack high-resolution
morphology to confirm signatures of interaction in these galaxies.  An
inhomogeneous intracluster gas distribution \citep{def03} might also
increase the frequency of triggering of star-forming activity in the
galaxies in the cluster central region; a direct confirmation of this
additional hypothesis would need a larger inventory of morphology and
spectra for the galaxies spatially coincident with a deep X-ray
emission map.  On the other hand, the archival \HST\ morphology
suggests galaxy-galaxy harassment is very effective in inducing a
starburst even in the galaxies residing within the cluster central
region.  Galaxy harassment is also preferred over major mergers
because of the assumed preservation of disk structures.

\acknowledgments

We are especially grateful to Tadayuki Kodama for sharing with us
their photometric redshifts catalog of \object{Abell 851}, without
which the detailed study of our sample in terms of local environments
could not be carried out.  We are also thankful to Betty De Filippis
(for the X-ray contour map) and Francois Hammer and Hector Flores (for
the \CFRS\ spectra) for giving us an access to their data.  We also
thank the anonymous referee for numerous valuable comments which
helped us make this a stronger paper.  T.S. thanks Shane Davis,
Colleen Schwartz, and Tommaso Treu for useful conversations on various
topics which constituted portions of this paper.
% Formal
Financial support was provided by the David and Lucille Packard
Foundation and the Alfred P. Sloan Foundation.  This research has made
use of the NASA/IPAC Extragalactic Database (NED), which is operated
by the Jet Propulsion Laboratory, California Institute of Technology,
under contract with the National Aeronautics and Space Administration.
This research has made use of the NASA Astrophysics Data System
abstract service.  Funding for the DEEP2 survey has been provided by
NSF grant AST-0071048 and AST-0071198.   Funding for the Sloan
Digital Sky Survey (SDSS) has been provided by the Alfred P. Sloan
Foundation, the Participating Institutions, the National Aeronautics
and Space Administration, the National Science Foundation, the
U.S. Department of Energy, the Japanese Monbukagakusho, and the Max
Planck Society. The SDSS Web site is http://www.sdss.org/.  The SDSS
is managed by the Astrophysical Research Consortium (ARC) for the
Participating Institutions. The Participating Institutions are The
University of Chicago, Fermilab, the Institute for Advanced Study, the
Japan Participation Group, The Johns Hopkins University, the Korean
Scientist Group, Los Alamos National Laboratory, the
Max-Planck-Institute for Astronomy (MPIA), the Max-Planck-Institute
for Astrophysics (MPA), New Mexico State University, University of
Pittsburgh, University of Portsmouth, Princeton University, the United
States Naval Observatory, and the University of Washington.
% For Keck
The authors wish to recognize and acknowledge the very significant
cultural role and reverence that the summit of Mauna Kea has always
had within the indigenous Hawaiian community.  We are most fortunate
to have the opportunity to conduct observations from this mountain.

{\it Facilities:} \facility{Keck:I (LRIS)}, \facility{Keck:II (DEIMOS)}

\appendix
\section{Spectral Evolution Modeling}
\label{sec.SpectralEvolutionModeling}

We synthesized model galaxy spectra incorporating differential
extinction of stellar continuum and nebular emission components.  We
used a combination of a stellar population synthesis code, GALAXEV
version 2003 \citep{bru03}, and a nebular photoionization code,
version 96.01a of Cloudy \citep{fer98}.  First, for each predefined
SFH, GALAXEV was used to synthesize a series of spectra at various
ages.  Three SFHs are explored: instantaneous burst; constant burst of
finite duration $\tau$, after which there is no star formation (i.e.,
truncation models); and exponentially-declining SFHs with various
$\tau$ (no gas recycling).  The analytical forms of these SFHs are
described in \citet{bru03}.  For the latter two SFHs, we used the
values of $\tau = 10$, $10^2$, $10^3$, and $2\times10^3~\Myr$.  We
used the GALAXEV models with the Padova~1994 evolutionary tracks and
assumed a Salpeter initial mass function ($\scM =
0.1$--$100\scM_\sun$) at solar abundance ($X=0.70$, $Y=0.28$, and
$Z=0.02$).  The standard GALAXEV output gives the production rates of
hydrogen-ionizing photons, $Q(\Hydrogen)$, which are scaled to the
total mass of an ionizing source ($\log{\scM_*/\scM_\sun} \sim
2$--$9$).  The mass of ionizing source therefore enters only as a
scaling factor for $Q(\Hydrogen)$.  Thus constructed ionizing source
spectrum energy distribution and properly scaled $Q(\Hydrogen)$ were
fed to Cloudy for photoionization simulations.

In Cloudy, our model \HII\ region is a single hydrogen-ionizing source
surrounded by a spherical distribution of constant-density, ionization
bounded nebula at the solar abundance.  We used the solar abundance
and depletion factors assumed by \citet{kew01}.  The hydrogen density
($n_{\rm H} = 10~\cm^{-3}$), the volume filling factor ($\epsilon =
1$), the covering factor ($C=1$), and the inner radius ($r_{\rm in} =
10^{-2}~\pc$) to the nebula were all kept constant throughout the
simulations with Cloudy.  These parameters can be chosen rather
arbitrary, since the ionization state of nebula is uniquely determined
by the combination of the spectral energy distribution of ionizing
source and the ionization parameter defined as
\[
u = \frac{Q(\Hydrogen)}{4\pi R_s^2 n_{\rm H} c} \ ,
\]
where $R_s$ is the \Stromgren\ radius and $c$ is the speed of light,
so long as the density remains below the limit for collisionally
de-excitation cooling \citep[see, e.g.,][]{sta96}.

Dust modeling features were generally turned off within these codes.
We manually attenuated the stellar and nebular components by a
different amount of extinction, adopting the attenuation curve of
\citet[][Eq.~(8)]{cal00}.  It has been shown empirically that the
stellar continuum suffers roughly half of the extinction suffered by
the nebular gas, $\EBV_* \simeq 0.44 \EBV_{\rm g}$ \citep[see][and
reference therein]{cal94}.  This ratio was varied [$\EBV_* / \EBV_{\rm
g} \equiv \EBVr = 0.22$, $0.44$, and $0.88$] to see the effect of
differential extinction on our measurements of equivalent widths.
This is motivated by the interpretation of e(a) spectrum by
\citet{pog01}, requiring selective extinction of light from stellar
populations of various ages.  A range of extinction $\EBV =
0.0$--$1.5$ was explored.  Note that $\EBV \equiv \EBV_{\rm g}$ is the
amount of extinction in the nebular component, being derived from a
Balmer decrement \citepalias{sat06}.  Resulting synthesized spectra
are used to measure \fOII\ and \Hd\ equivalent widths using the
flux-summing method \citepalias{sat06}.  The advantage of our method
presented here is that both emission and absorption lines, each
suffering a different amount of extinction, are modeled into the final
spectra, which are measured with the same tool used for measuring
other spectra.

\begin{deluxetable}{cccc}
\tabletypesize{\footnotesize}
\tablecaption{Comparison of Coupled-Model Simulations}
\tablewidth{0pt}
\tablehead{
\colhead{} &
\multicolumn{3}{c}{Models} \\
\cline{2-4} \\
\colhead{Line} &
\colhead{This Paper\tablenotemark{a}} &
\colhead{M01\tablenotemark{b}} &
\colhead{G95\tablenotemark{c}}
}
\startdata
$\log\Hb~[\erg~\s^{-1}]$ & 38.79 & 38.83 & 38.84 \\
$\fOII\lam3727/\Hb$   & 3.12 & 3.10 & 3.07 \\
$\fOII\lam5007/\Hb$   & 0.32 & 0.31 & 0.25 \\
$\fOI\lam6300/\Hb$    & 0.07 & 0.04 & 0.04 \\
$\fNII\lam6584/\Hb$   & 1.29 & 1.34 & 1.34 \\
$\fSII\lam6716/\Hb$   & 0.70 & 0.57 & 0.58 \\
$\fSII\lam6731/\Hb$   & 0.49 & 0.39 & 0.40 \\
$\fSIII\lam9069/\Hb$  & 0.31 & 0.36 & 0.34
\enddata

~\tablenotetext{a}{GALAXEV (version 2003) \& Cloudy (version 96.01a).}

~\tablenotetext{b}{PEGASE.2 \& Cloudy (version 90.04) by
\citet{moy01}.}

~\tablenotetext{c}{Cloudy (version unknown) and see \citet{gar95} for
their stellar population synthesis models. }

~\tablecomments{ The comparison of $Z=Z_\sun$ instantaneous burst
model at $1~\Myr$ and $\log u \simeq -3.11$.  The solar abundance is
as defined in \citet{gar95}. }

\label{tab.SFHSimModelComp}
\end{deluxetable}

We checked our method of simulation by running a simple model that
closely resembles the solar abundance, $M_T = 2\times10^4\scM_\sun$
model published by \citet[][Table~16]{gar95}, which has also been used
by \citet{moy01} to check their simulation for consistency.
\tab{tab.SFHSimModelComp} shows that the results generally agree well
with each other.

\begin{figure}
\epsscale{.5}
\plotone{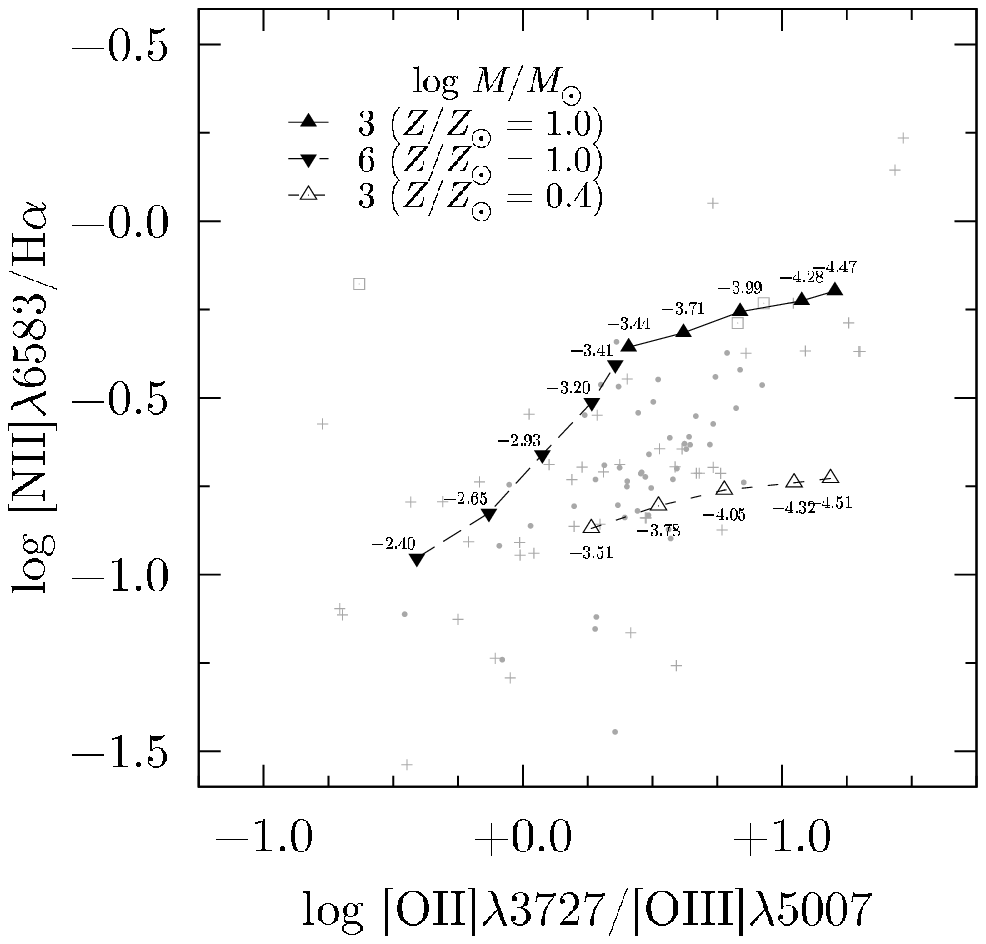}
\caption{ Emission line flux ratio $\fNII\lam6583/\Ha$
vs.~$\fOII\lam3727/\fOIII\lam5007$ for three evolutionary model
tracks.  The numbers along each track indicate the ionization
parameter $u$ for each model.  For each evolutionary track, data
points appear at $1$, $10$, $10^2$, $10^3$, and $5\times10^3~\Myr$
from high to low $u$.  Gray data points indicate the cluster \fOII\
emitters whose line ratios are computed from their highest-\snr\
spectrum (filled dots) and secondary spectrum (crosses).  Squares
indicate AGN-like galaxies.  See \citetalias{sat06} for the definition
of highest-\snr\ spectrum, secondary spectrum, and AGN-like galaxies;
see \citetalias{sat06}.  }
\label{fig.LogN2HaVsLogO2O3}
\end{figure}

Although our interest mainly lies in the evolution of spectral indices
in the \fOII-\Hd\ plane, we need some constraint on the relevant range
of ionization parameter $u$ so that our models roughly resemble the
ionization state of nebulae in the observed galaxies.  In our models,
the ionization parameter is essentially a function only of the
production rate of hydrogen-ionizing photons $Q(\Hydrogen)$, and
electron temperature $T_{\rm e}$, namely
\[
u = A(T_{\rm e}) \left[ Q(\Hydrogen) n_{\rm H} \epsilon^2
\right]^{1/3} \ ,
\]
where $A(T_{\rm e}) \approx 2.8\times10^{-20}(10^4/T_{\rm e})^{2/3}$
\citep{sta96} and the $T_{\rm e}$ dependence comes from $R_s$ (which
depends on the recombination coefficient).  Since the temperature
varies within a nebula, we simply take the value of $u$ computed by
Cloudy at the end of each simulation.  Hence adjusting $u$ is done
through one free parameter, the mass of central ionizing source with
which $Q(\Hydrogen)$ scales proportionally.  Since the strength of
$\fOII\lam3727$ emission depends on the relative abundance of
different oxygen ionic species, we constrained this by comparing the
ratio $\fOIII\lam5007/\fOII\lam3727$ between the \fOII\ emitting
sample and our simulated models.  In \fig{fig.LogN2HaVsLogO2O3} we see
how solar-abundance models compare to observed galaxies.  We first
notice that any single model track between the ages
$t=(1$--$5)\times10^3~\Myr$ does not span the range of observed
$\fOII/\fOIII$ entirely, suggesting a need for variety of SFHs to
explain the wide range of observed line ratios.  Moreover, our models
yield generally higher $\fNII\lam6583/\Ha$.  The ratio $\fNII/\Ha$ is
sensitive both to the relative abundance of N to all other species and
to the overall metal abundance of nebula, but it has relatively little
effect on the observables in the \fOII-\Hd\ plane.  Although the line
ratio diagram implies the cluster \fOII\ emitters may be at subsolar
abundance in general, we chose solar-metallicity, $\scM = 4\scM_\sun$
models at $\EBV = 0.5$ and $\EBVr = 0.44$ as our fiducial set of
models.

\end{document}